# Beyond the boundaries of the second law of thermodynamics: Intrinsic-nonequilibrium distribution of large ions in charged small nanopores

Yu Qiao,[1,2,*] Meng Wang[2]

[1] *Program of Materials Science and Engineering, University of California – San Diego, La Jolla, CA 92093, U.S.A.*
[2] *Department of Structural Engineering, University of California – San Diego, La Jolla, CA 92093-0085, U.S.A.*
[*] Email: yqiao@ucsd.edu

**Abstract.** Recent theoretical research on the fundamentals of statistical mechanics has led to a counterintuitive finding [2-5]: with a locally nonchaotic energy barrier, a macroscopic many-particle system may reach an intrinsic-nonequilibrium steady state without any external driving force, incompatible with the conventional statements of the second law of thermodynamics. In the present investigation, we experimentally demonstrate the concept of intrinsic nonequilibrium by replacing the weak gravitational force in the "toy model" with a strong Coulomb force. The tests are performed on a set of nanoporous carbon electrodes immersed in aqueous cesium pivalate solutions. The key characteristic is that the effective nanopore size is only slightly larger than the effective ion size but less than twice the ion size, so that the motion of the confined ions is quasi-one-dimensional and relatively nonchaotic. At first glance, the supercapacitive cells exhibit "normal" charge curves. However, the steady-state distribution of the large ions in the charged small nanopores (measured by the charge efficiency Λ) inherently differs from equilibrium by nearly one order of magnitude, thereby breaking the boundaries of the second law of thermodynamics and potentially opening the possibility of producing useful work by absorbing heat from a single thermal reservoir with no other effect. Such an unusual phenomenon is compatible with the basic principle of maximum entropy and does not challenge the second law within its established domain of application (fully chaotic systems); instead, it extends our understanding of thermodynamics beyond the traditional framework.

## 1. Introduction

1.1 Nonchaoticity and the boundaries of thermodynamics

The second law of thermodynamics is a fundamental principle of physics, yet its boundaries have not been fully clarified. A key issue in establishing the second law is time reversibility [1]: as the microscopic equations of motion (e.g., Newton's laws or the Schrödinger equation) are time-reversible, how can entropy increase irreversibly in an isolated system? Historically, the most influential attempt to resolve this problem is Boltzmann's H-theorem [1], which is based on the hypothesis of molecular chaos: before a random particle-particle collision, the two-particle probability density is replaced by the product of the one-particle probability densities, thereby rendering the probability of system evolution asymmetric in time. However, this framework may not apply if the particle-particle interactions are not extensive, either throughout the system or in a local region [2-5].

In this manuscript, the central question is whether a system can exist that simultaneously satisfies the following three conditions: (1) it is macroscopic in scale and consists of many particles; (2) it cannot relax to thermodynamic equilibrium when immersed in a thermal reservoir; (3) the energetic consequences are nontrivial. We refer to such a system as exhibiting *intrinsic* nonequilibrium, defined as a steady state that differs significantly from equilibrium in the absence of external thermodynamic driving forces (see Section A1 of the Appendix for further discussion). An intrinsic-nonequilibrium particle distribution refers to a steady state in which, between two energy states, the ratio of particle number densities is inherently different from the Boltzmann factor $e^{-\beta\cdot\Delta V_{\mathrm{z}}}$, where $\Delta V_{\mathrm{z}}$ is the potential difference, $\beta = 1/(k_{\mathrm{B}}T)$, $k_{\mathrm{B}}$ is the Boltzmann constant, and $T$ is temperature. We do not study transient or fluctuating processes.

On the one hand, for a fully chaotic system, without an energetic penalty, the second law of thermodynamics forbids intrinsic nonequilibrium [1]. For instance, across a porous membrane in an isothermal or isolated ideal gas, the pressure must be uniform, regardless of the pore size or the pore geometry; a nonuniform steady-state gas distribution would cause unusual effects [5]. The thought experiment of Maxwell's demon attempts to interrupt the equilibrium state, which has inspired exploration of the physical nature of information [e.g., 6,7].

On the other hand, it is well known that small-sized intrinsic-nonequilibrium setups exist, such as certain nonchaotic or nonergodic particle movements [8]. One example is a Knudsen gas [4], i.e., a rarefied gas with the Knudsen number ($K_{\mathrm{g}} = \lambda_{\mathrm{F}}/d_{\mathrm{c}}$) larger than 1, where $\lambda_{\mathrm{F}}$ is the mean free path of the gas particles and $d_{\mathrm{c}}$ is the gas-body size (e.g., the characteristic length scale of the gas container). The gas container is immersed in a thermal reservoir. As particle-particle collisions rarely occur, faster gas particles tend to collide more rapidly with the container walls and release heat to the environment (the thermal reservoir), while slower gas particles tend to stay longer in the interior. Consequently, the effective gas-phase kinetic temperature is lower than the thermal-reservoir temperature. Traditionally, such intrinsic-nonequilibrium phenomena were not viewed as thermodynamic problems, because the systems are small and their energetic effects are "trivial," i.e., they can produce little useful work.

Recently, we investigated a concept that extends beyond the boundaries of the second law of thermodynamics [2-4]: the spontaneous-nonequilibrium domain (SND), sometimes also referred to as the local area spontaneously out of equilibrium or the spontaneously nonequilibrium dimension. In general, when the system is at rest (i.e., without any external thermodynamic driving force), if the distribution of particle number density or kinetic temperature across a local region (often narrower than the particle mean free path $\lambda_{\mathrm{F}}$) cannot relax to equilibrium, that region is defined as a SND.

The analysis of a set of quantum models [3] suggests that SND favors localized wave packets and unquantized energy, having a tendency to be a semiclassical or classical-mechanical phenomenon. Sections 1.2-1.4 below briefly introduce the "toy models" studied in [2].

### 1.2 Knudsen gas in a gravitational field

Figure 1(a) depicts a two-dimensional (2D) vertical plane, wherein a number of 2D elastic particles (finite-sized hard disks) randomly move [2]. The plane height is denoted by $z_{\mathrm{G}}$. The system is closed and immersed in a thermal reservoir, in a uniform gravitational field ($g$). The top and bottom boundaries are thermal walls at constant $T$, and the left and right boundaries are periodic. When a particle collides with a thermal wall, the reflected speed is uncorrelated with the incident speed but randomly follows the Maxwell-Boltzmann distribution at $T$; the reflected direction is random. The details of this boundary setting are nonessential to the rest of the discussion. Various other boundary conditions have been examined in [2-4].

If $z_{\mathrm{G}} \gg \lambda_{\mathrm{F}}$, the particle-particle collisions between the top and bottom walls are extensive. As entropy ($S$) is maximized, the particle flux ratio ($\delta \triangleq n_{\mathrm{t}}/n_{\mathrm{b}}$) is the Boltzmann factor $\delta_0 = e^{-\beta m g z_{\mathrm{G}}}$, where $m$ is the particle mass, and $n_{\mathrm{t}}$ and $n_{\mathrm{b}}$ are the numbers of particle-wall collisions at the top and bottom boundaries, respectively.

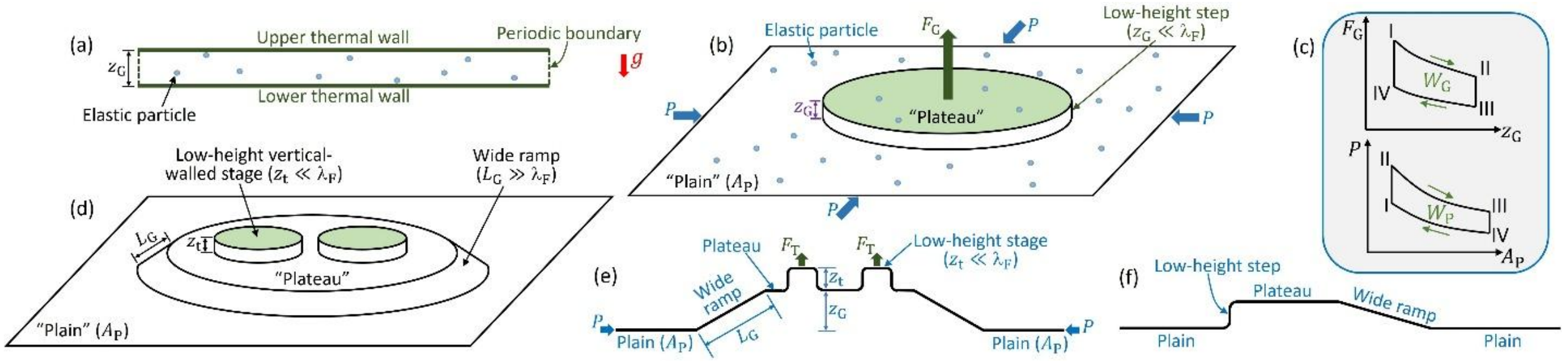

**Figure 1.** A few locally nonchaotic models that break the boundaries of the second law of thermodynamics [2]. They are fundamentally different from Feynman's ratchet or Maxwell's demon. **(a)** A Knudsen gas in a gravitational field ($g$). The left and right boundaries are open and periodic. The top and bottom boundaries are thermal walls, at which the reflected particle speed randomly follows the Maxwell-Boltzmann distribution, uncorrelated with the incident speed. When the system height ($z_{\mathrm{G}}$) is much less than the nominal particle mean free path ($\lambda_{\mathrm{F}}$), the particle-particle collisions are sparse, and the distribution of particle number density along the vertical z-dimension is inherently non-Boltzmannian. **(b)** The two-shelf ideal-gas model. The system is closed and immersed in a thermal reservoir. A large number of elastic particles randomly move on the upper shelf (the "plateau") and the lower shelf (the "plain") across the low-height step, in a gravitational field ($g$). The step height ($z_{\mathrm{G}}$) is much less than $\lambda_{\mathrm{F}}$, so that the particle trajectories in the step tend to be independent of each other, i.e., the step is a locally nonchaotic energy barrier (the SND). A remarkable consequence of $z_{\mathrm{G}} \ll \lambda_{\mathrm{F}}$ is that the plateau-to-plain ratio of particle number density ($\bar{\rho}_{\mathrm{G}}$) is significantly smaller than the Boltzmann factor $\delta_0 = e^{-\beta m g z_{\mathrm{G}}}$. **(c)** The four-step isothermal cycle. The plateau height ($z_{\mathrm{G}}$) and the plain area ($A_{\mathrm{P}}$) are alternately adjusted by the support force ($F_{\mathrm{G}}$) and the in-plane pressure ($P$), respectively. Indexes I-IV indicate the system states. Only when the system is in equilibrium (i.e., $\bar{\rho}_{\mathrm{G}} = \delta_0$), can the total produced work ($W_{\mathrm{P}}$) be equal to the total consumed work ($W_{\mathrm{G}}$). With the low-height step, because $\bar{\rho}_{\mathrm{G}} < \delta_0$, $W_{\mathrm{P}}$ is significantly larger than $W_{\mathrm{G}}$, incompatible with the conventional heat-engine statement of the second law. **(d)** Elevated view and **(e)** side view of a variant of the model in panel (b). The elastic particles are not shown. The plateau-plain boundary is a wide ramp. The ramp size $L_{\mathrm{G}} \gg \lambda_{\mathrm{F}}$. There are low-height vertical-walled stages distributed on the plateau. The stage height $z_{\mathrm{t}} \ll \lambda_{\mathrm{F}}$. The stage walls act as locally nonchaotic energy barriers (the SND). **(f)** Side view of another variant model. The elastic particles are not shown. A net particle flow can be spontaneously generated from the random thermal motion of the particles, without any energetic penalty.

If $z_{\mathrm{G}} \ll \lambda_{\mathrm{F}}$, the system is a Knudsen gas. Without extensive particle-particle collisions, the particle trajectories tend to be independent of each other. Whether a particle can overcome the gravitational energy barrier and ascend from the bottom wall to the top wall is determined by its kinetic energy associated with vertical motion ($K_\mathrm{z}$), relatively unrelated to the horizontal motion. Since the average $K_\mathrm{z}$ is only half of the average particle kinetic energy ($\bar{K} = k_\mathrm{B}T$), the particle flux ratio ($\delta$) is smaller than $\delta_0$.

When particle-particle collisions are negligible, the intrinsic-nonequilibrium particle-flux ratio ($\delta \triangleq n_\mathrm{t}/n_\mathrm{b}$) may be calculated as

$$\delta_\mathrm{k} = \frac{\int_{\sqrt{2gz_\mathrm{G}}}^{\infty} \frac{p_\mathrm{z}(v_\mathrm{z})}{t_\mathrm{r}(v_\mathrm{z})} \mathrm{d}v_\mathrm{z}}{\int_0^{\infty} \frac{p_\mathrm{z}(v_\mathrm{z})}{t_\mathrm{r}(v_\mathrm{z})} \mathrm{d}v_\mathrm{z}}, \quad (1)$$

where $v_\mathrm{z}$ denotes the component of the particle velocity along the vertical z-axis; $p_\mathrm{z}(v_\mathrm{z}) = \sqrt{2\beta m/\pi}\, e^{-\beta m v_\mathrm{z}^2/2}$ is the one-dimensional (1D) Maxwell-Boltzmann distribution of $|v_\mathrm{z}|$; the round-trip time $t_\mathrm{r}(v_\mathrm{z})$ is $2v_\mathrm{z}/g$ (for $v_\mathrm{z} < \sqrt{2gz_\mathrm{G}}$) or $t_\mathrm{u} + \bar{t}_\mathrm{d}$ (for $v_\mathrm{z} \geq \sqrt{2gz_\mathrm{G}}$), with the effective ascending and descending transit times being $t_\mathrm{u} = \left(v_\mathrm{z} - \sqrt{v_\mathrm{z}^2 - 2gz_\mathrm{G}}\right)/g$ and $\bar{t}_\mathrm{d} = \left(\sqrt{k_\mathrm{B}T/m + 2gz_\mathrm{G}} - \sqrt{k_\mathrm{B}T/m}\right)/g$, respectively. In [2], as a first-order approximation, the effects of $t_\mathrm{r}$ are neglected and $\delta_\mathrm{k}$ is simplified to

$$\delta_\mathrm{k} \approx \delta_1 \triangleq \int_{\sqrt{2gz_\mathrm{G}}}^{\infty} p_\mathrm{z}(v_\mathrm{z}) \mathrm{d}v_\mathrm{z} = 1 - \mathrm{erf}(\sqrt{\beta m g z_\mathrm{G}}). \quad (2)$$

Approximating $\delta_\mathrm{k}$ by $\delta_1$ is done for analytical convenience and does not affect the numerical simulations in [2]. This approximation ($\delta_k \approx \delta_1$) is conservative, since $1/t_\mathrm{r}$ tends to assign greater weight to smaller velocities when $v_\mathrm{z} < \sqrt{2gz_\mathrm{G}}$. In the parameter ranges considered in [2], $\delta_1$ is about 10% larger than $\delta_\mathrm{k}$, and both $\delta_\mathrm{k}$ and $\delta_1$ are nearly a factor of two smaller than $\delta_0$. The details of the $v_\mathrm{z} - z$ relationship have been further studied in [4].

The significant difference between $\delta_0$ and $\delta_\mathrm{k}$ (or $\delta_1$) clearly indicates that the particle number density distribution is inherently non-Boltzmannian ($\delta \neq \delta_0$). That is, the system cannot relax to equilibrium. By itself, $\delta_\mathrm{k} < \delta_0$ (or $\delta_1 < \delta_0$) is unsurprising, as a Knudsen gas is known to be non-thermodynamic.

It is possible to use $p_\mathrm{z}(v_\mathrm{z})$ to construct integrals that are equal to the Boltzmann factor $\delta_0$, e.g., $\bar{\rho}_\mathrm{k} = \int_{\sqrt{2gz_\mathrm{G}}}^{\infty} [\bar{\alpha}_\mathrm{m} v_\mathrm{z} \cdot p_\mathrm{z}(v_\mathrm{z})] \mathrm{d}v_\mathrm{z}$, where $\bar{\alpha}_\mathrm{m} = \sqrt{(\pi\beta m/2)^{1/2}}$. However, for the discussion of the crossing ratio $\delta$, in the integrand of $\bar{\rho}_\mathrm{k}$, neither the coefficient $\bar{\alpha}_\mathrm{m}$ nor the factor $v_\mathrm{z}$ has a clear physical meaning. In fact, $\bar{\rho}_\mathrm{k} = \delta_0$ supports that $\delta_\mathrm{k} \neq \delta_0$. As demonstrated in [2], since the expectation value of $v_z$ depends on height $z$, $\delta$ is necessarily in an out-of-nonequilibrium state (i.e., $\delta \neq \delta_0$).

From a different perspective, if $\delta_K$ were equal to $\delta_0$, it would imply that the second law of thermodynamics could be derived from Newton's second law by analyzing an individual particle, which contradicts fundamental concepts of statistical mechanics, such as the H-theorem [1]. It cannot explain why the increase of entropy of an isolated system is irreversible, while the equations of motion are time-reversible.

1.3 Locally nonchaotic two-shelf ideal-gas model

When a Knudsen-gas-type component is incorporated into a large system, nontrivial phenomena may arise. Figure 1(b) depicts a two-shelf ideal-gas model [2,9]. In a uniform gravitational field ($g$), a large number of elastic particles randomly move on the upper shelf (the "plateau") and the lower shelf (the "plain") across the low-height step. The two shelves are perfectly horizontal. The system is closed and immersed in a thermal reservoir at $T$. The plateau can be raised or lowered by the support force $F_G$. The plain can be expanded or compressed by the in-plane pressure $P$.

When the plateau height ($z_G$) is much less than $\lambda_F$, the step is the SND. Similar to the vertical plane in Figure 1(a), inside the nonchaotic step, the particles tend to ascend or descend individually; across the step, the particle flux ratio $\delta \approx \delta_1$. A remarkable consequence is that the steady-state plateau-to-plain ratio of particle number density is not in equilibrium: $\bar{\rho}_G \triangleq \rho_G/\rho_P \approx \delta_1 \neq \delta_0$, where $\rho_G$ and $\rho_P$ are the particle number densities on the plateau and the plain, respectively. As particle-particle collisions are sparse in the step, there is no driving force for the system to reach thermodynamic equilibrium.

Under the condition of $\bar{\rho}_G \neq \delta_0$, an isothermal cycle can be designed to produce useful work by absorbing heat from the environment (a single thermal reservoir) with no other effect. Figure 1(c) shows the 4-step operation cycle, wherein $F_G$ and $P$ operate alternately. From State I to II, $z_G$ is increased by $F_G$, and $A_P$ is kept constant. From State II to III, $z_G$ is unchanged, and $A_P$ is expanded by $P$. From State III to IV, $A_P$ is fixed, and $z_G$ is reduced back to the original height. Finally, from State IV to I, $A_P$ is compressed, and the system returns to the initial state.

As explained in Section A2 of the Appendix, the second law of thermodynamics can be formulated as the generalized Maxwell's relations [2]

$$\frac{\partial F_1}{\partial x_2} = \frac{\partial F_2}{\partial x_1}, \tag{3}$$

where $F_1$ and $F_2$ are two thermally correlated thermodynamic forces, and $x_1$ and $x_2$ are their conjugate variables, respectively. For an equilibrium system, Equation (3) can be derived from $F_1 = \partial F_\mathrm{A}/\partial x_1$ and $F_2 = \partial F_\mathrm{A}/\partial x_2$, with $F_\mathrm{A}$ being the Helmholtz free energy. When the plateau area ($A_\mathrm{G}$) and the plain area ($A_\mathrm{P}$) are much larger than the step area, $F_\mathrm{G} = mgN_\mathrm{G}$ and $PA_\mathrm{P} = N_\mathrm{P}k_\mathrm{B}T$ [9], with $N_\mathrm{G}$ and $N_\mathrm{P}$ being the particle numbers on the plateau and in the plain, respectively. Thus, for $F_\mathrm{G}$ and $P$, Equation (3) becomes $-\partial F_\mathrm{G}/\partial A_\mathrm{P} = \partial P/\partial z_\mathrm{G}$, which may be rewritten as $\partial \bar{\rho}_\mathrm{G}/\partial z_\mathrm{G} = -\beta mg$ and its solution is $\bar{\rho}_\mathrm{G} = e^{-\beta mg z_\mathrm{G}}$. That is, in the isothermal cycle in Figure 1(c), Equation (3) is satisfied only when the system is in equilibrium ($\bar{\rho}_\mathrm{G} = \delta_0$), so that the work produced by the in-plane pressure $P$ ($W_\mathrm{P}$) is equal to the work consumed by the support force $F_\mathrm{G}$ ($W_\mathrm{G}$). With the locally nonchaotic SND (the low-height step), because $\bar{\rho}_\mathrm{G} < \delta_0$, $W_\mathrm{P}$ is considerably larger than $W_\mathrm{G}$. Hence, useful work is produced in each cycle, incompatible with the conventional heat-engine statement of the second law.

## 1.4 Variant models

Figure 1(d,e) depicts a variant of the model in Figure 1(b) [2]. The plateau-plain boundary is a wide ramp, with the ramp size ($L_\mathrm{G}$) much larger than $\lambda_\mathrm{F}$. A number of vertical-walled stages are distributed on the plateau. The stage floors are connected to the plateau, keeping the stage height ($z_\mathrm{t}$) proportional to the plateau height ($z_\mathrm{G}$). When $z_\mathrm{t} \ll \lambda_\mathrm{F}$, the stage walls act as the SND, rendering the particle number density distribution out of equilibrium. Specifically, the number of particles on the stages ($N_\mathrm{T}$) and the associated support force ($F_\mathrm{T} \propto mgN_\mathrm{T}$) do not follow the Boltzmann factor $e^{-\beta mgh_\mathrm{t}}$, where $h_\mathrm{t} = z_\mathrm{G} + z_\mathrm{t}$; $F_\mathrm{T}$ is a part of the total support force ($F_\mathrm{G}$) that simultaneously adjusts the plateau height and the stage height. Like Figure 1(b), Figure 1(d,e) does not satisfy Equation (3). In the isothermal cycle in Figure 1(c), $W_\mathrm{P} > W_\mathrm{G}$. The produced work ($\Delta W = W_\mathrm{P} - W_\mathrm{G}$) is from heat absorbed from the environment.

In Figure 1(b,d), fundamentally different from Maxwell's demon, the out-of-equilibrium steady state is rooted in the intrinsic nonchaotic nature of the SND and does not involve any extrinsic thermodynamic driving force or information processing. There are various other ways to arrange the SND. For instance, Figure 1(f) shows an asymmetric plateau-plain setup [2]. Unlike Feynman's ratchet, the plateau height ($z_{\mathrm{G}}$) is much less than $\lambda_{\mathrm{F}}$. One end of the plateau is connected to the plain through a low-height step (the SND), and the other end is connected through a wide ramp. The ramp size $L_{\mathrm{G}}$ is much larger than $\lambda_{\mathrm{F}}$. A large number of particles randomly move on the plateau and the plain across the step and the ramp. Because the particle movement in the ramp is chaotic, across it the local particle flux ratio $\delta \rightarrow \delta_0$. Yet, as discussed for Figure 1(a), across the low-height step, $\delta \approx \delta_1$. Since $\delta_1 < \delta_0$, the overall probability for the particles to move from the plain to the plateau across the ramp is higher than that of the step. At the steady state, a net particle flow can be generated. It leads to entropy decrease without an energetic penalty, incompatible with the conventional entropy statement of the second law of thermodynamics.

Besides the intrinsic-nonequilibrium distribution of particle number density, SND can also result in unusual thermal states [4]. In Figure 1(a), when $z_{\mathrm{G}} \ll \lambda_{\mathrm{F}}$ and the particle trajectories are nonchaotic, thermal equilibrium cannot be reached. The average particle kinetic energy ($\overline{K}$) varies with height $z$. Consequently, heat may be spontaneously and continuously transported by the gas particles from the cold side to the hot side, incompatible with the conventional refrigeration statement of the second law of thermodynamics.

## 1.5 Challenges to the experimental study

In a thermodynamic system, the presence of SND is not allowed by the second law. In Figure 1(b), compared to other narrow bands, the low-height step is special in the sense that gravity acts within it but not outside it. With this configuration, as the local particle behavior influences the global state, the non-Boltzmann characteristics traditionally unique to small setups "spread" from the narrow SND to the large shelves, causing the global intrinsic-nonequilibrium state. SND is beyond the scope of Boltzmann's H-theorem, since the hypothesis of molecular chaos does not account for local nonchaoticity.

While the SND models in Figure 1 have interesting properties, directly testing them is difficult, primarily because of the extremely large $g$ required. To achieve a substantial intrinsic-nonequilibrium effect, if the particles are ambient air molecules, $g$ needs to be higher than $10^{11}$ m/s$^2$, at the level of neutron stars.

In this paper, the concept of Figure 1(d,e) is experimentally investigated, with the weak gravitational force being replaced by a strong Coulomb force. The working medium is changed from a rarefied gas to a dilute electrolyte solution. Section 2 presents the key component of the experimental design: large ions in charged small nanopores. Section 3 presents the thermodynamic analysis of the experimental setup. The second law of thermodynamics dictates that the steady state must be in equilibrium, i.e., the effective surface ion density must be proportional to the Boltzmann factor. Sections 4-5 present the experimental procedure and results: remarkably, the measured steady-state ion distribution is highly non-Boltzmannian.

## 2. Key component in the experiment: Large ions in charged small nanopores

Figure 2(a) shows the key component in our experiment: a microporous electrode immersed in an electrolyte solution. Compared with Figure 1(d,e), when the electrode is charged, the high potential in the micropores corresponds to the upper plateau; changing the electric potential ($V$) is analogous to adjusting the plateau height. The bulk liquid phase is the counterpart of the lower plain; reducing the electrolyte concentration ($c$) is analogous to expanding the plain, and vice versa. The charge carriers (the dissolved ions) are the counterpart of the elastic particles. The equilibrium ion exchange between the outer electrode surface and the bulk liquid phase corresponds to the wide ramp. The energy barrier is formed by the Coulomb force, which is much stronger than the gravitational force, desirable for experimental study.

The system has two critical characteristics. First, the effective pore size ($d_{\mathrm{e}}$) must be only slightly larger than the effective ion size ($d_{\mathrm{i}}$), less than $2d_{\mathrm{i}}$. Under this condition, in a nanopore, the confined ions form a quasi-1D lineup. As demonstrated by the computer simulations in open literature [10-16] and as will be discussed in Section 6.2, the confinement effect of the nanopore walls may cause the steady-state ion distribution to be intrinsically out of equilibrium, comparable to the role of the low-height stage walls in Figure 1(d,e). Second, the electrolyte concentration in the bulk liquid phase ($c$) should be relatively low. When $c$ is less than 20 mM, with a constant electrode charge ($Q$), the cell potential ($V$) is sensitive to the variation in $c$.

People have long been aware that when $d_{\mathrm{e}}$ is slightly larger than $d_{\mathrm{i}}$, the behavior of large ions in charged small micropores may be unusual. A number of molecular dynamics (MD) simulations on supercapacitors [10-16] have consistently demonstrated that the effective ion diffusion coefficient in micropores ($D_{\mathrm{i}}$) can be highly dependent on the electric potential ($V$). For example, upon quasistatic charging of a 0.75-nm-diameter pore [10], the diffusion coefficients of 1-ethyl-3-methylimidazolium cations ($EMIM^{+}$) and bis(trifluoro-methylsulfonyl)-imide anions drastically increase by two orders of magnitude. As a result, the total surface ion density ($\sigma_{\mathrm{tot}} = \sigma^{+} + \sigma^{-}$) may decrease with $Q$, where $\sigma^{+}$ and $\sigma^{-}$ are the effective surface densities of counter-ions and co-ions, respectively. If the pore size is increased to over 1 nm, the correlation between $D_{\mathrm{i}}$ and $V$ abruptly vanishes. For another example [12], when $d_{\mathrm{e}}$ is 0.9 nm, as $V$ rises, the diffusion coefficients of $EMIM^{+}$ in both the cathode and the anode vary by one order of magnitude. In a 1-nm-diameter pore [15], depending on the ion-ion interaction, $\sigma_{\mathrm{tot}}$ may be lower at a higher $V$. The unusually fast ion diffusion is observed at steady state without any additional electrical energy consumption. The diffusion is thermally driven. As the adsorbed ions diffuse along the axial direction of a nanopore, the state of charge and electrode potential remain constant. Regardless of how long the system evolves, the effective surface ion density does not reach equilibrium.

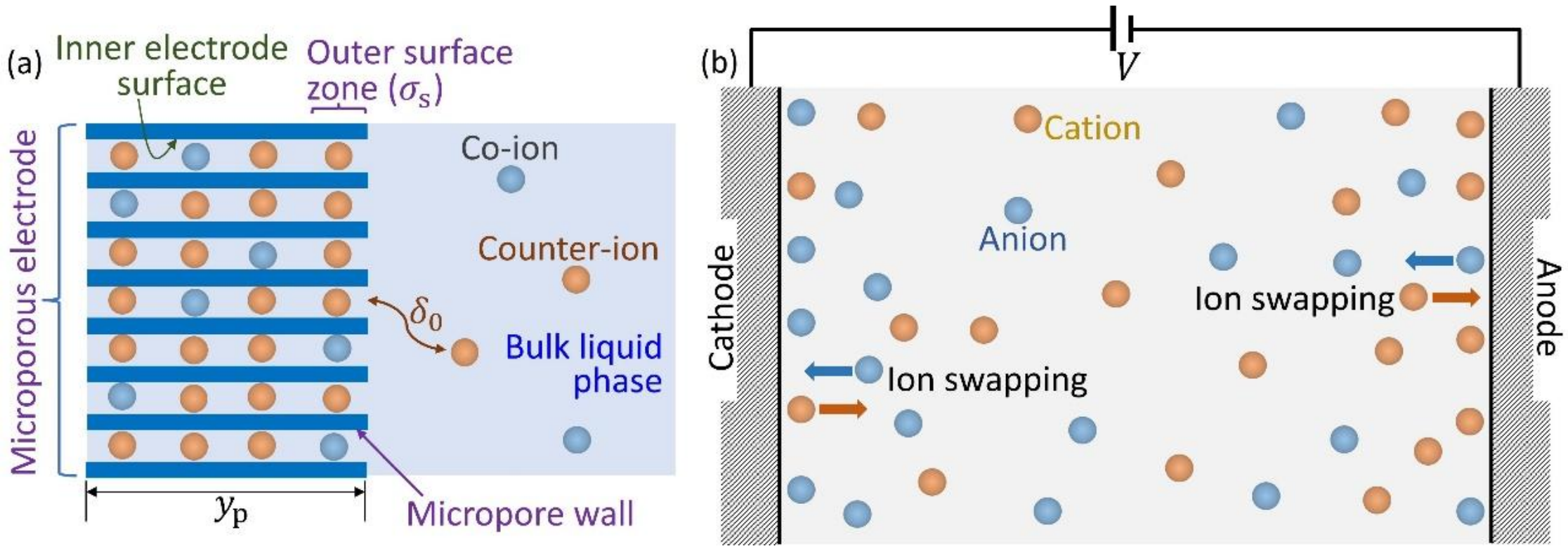

**Figure 2. (a)** A charged microporous electrode immersed in an electrolyte solution, with adsorbed co-ions and counter-ions. In a micropore, when the effective pore size is only slightly larger than the effective ion size, the confined ions form a quasi-1D lineup. Compared with Figure 1(d,e), the high-potential micropores are the counterpart of the plateau; the bulk liquid phase is analogous to the plain; the confinement effect of the nanopore walls plays a role somewhat similar to the locally nonchaotic low-height stage walls (the SND). **(b)** Schematic of the charging process of flat electrodes in an electrolyte solution. When the electrode charge ($Q$) is increased, the adsorption of counter-ions is accompanied by the repulsion of co-ions, as indicated by the arrows (the ion-swapping effect). Consequently, the increase in the number of ions adsorbed at the electrode surfaces (i.e., the loss of ions from the bulk liquid phase) is smaller than the increase in $Q/z$, where $z$ is the ionic charge. That is, the charge efficiency ($\Lambda$) is smaller than 1.

Since $D_{\mathrm{i}}$ depends on $V$, when the electrodes are charged, the steady-state ion distribution is not in equilibrium, i.e., $\sigma^{+}$ and $\sigma^{-}$ are not proportional to the Boltzmann factor $\delta_0 = e^{\mp \beta z e_0 V/2}$; hereafter, $z$ denotes the ionic charge and $e_0$ denotes the elementary charge. As depicted in Figure 2(a), the number of adsorbed ions in the electrode can be written as $N_{\mathrm{ad}} \triangleq \sigma_{\mathrm{tot}} A_{\mathrm{e}} = N_{\mathrm{s}}(y_{\mathrm{p}}/\Delta\bar{y})$, where $A_{\mathrm{e}}$ is the effective surface area of the electrode, $N_{\mathrm{s}} \triangleq \sigma_{\mathrm{s}} A_{\mathrm{s}}$ is the number of ions in the outer surface zone of the electrode, $\sigma_{\mathrm{s}}$ is the effective surface-zone ion density, $A_{\mathrm{s}}$ is the outer surface area of the electrode, $y_{\mathrm{p}}$ is the effective micropore length, and $\Delta\bar{y}$ is the average ion spacing in the micropores. The outer surface zone can directly exchange ions with the bulk liquid phase. Thus, $\sigma_{\mathrm{s}}$ tends to be proportional to $c\delta_0$. Since $D_{\mathrm{i}}$ is a function of $V$, $\Delta\bar{y}$ varies with $V$, such that the $N_{\mathrm{ad}} - V$ relationship is in a non-Boltzmann form. A more detailed analysis will be given in Section 3, indicating that the intrinsic-nonequilibrium $\sigma^{+}$ and $\sigma^{-}$ are incompatible with the conventional heat-engine statement of the second law of thermodynamics.

## 3. Thermodynamic analysis

The thermodynamic analysis in this section reaffirms that, in a capacitive cell under isothermal conditions, the second law of thermodynamics forbids an intrinsic-nonequilibrium ion distribution at steady state; i.e., the relationship between $V$ and $\sigma^{\pm}$ must follow the Boltzmann factor ($\delta_0$). This requirement conflicts with the MD simulations in [10-16] and will be examined by the experiment in Sections 4-5.

### 3.1 Cross-influence between electric potential and chemical potential

The second law of thermodynamics can be formulated by applying Equation (3) to Figure 2(a). In a capacitive cell, there are two thermally correlated thermodynamic forces: the chemical potential of the bulk liquid phase ($\mu$) and the electric potential of the electrodes ($V$) [17]. They are counterparts of $P$ and $F_{\mathrm{G}}$ in Figure 1(d,e), respectively. The conjugate variable of $\mu$ is the number of dissolved ions in the bulk liquid phase, $N_{\mathrm{e}}$. The conjugate variable of $V$ is the electrode charge, $Q$. For an equilibrium system, in accordance with Equation (3), the cross-influence of $\mu$ and $V$ must be balanced [17,18], i.e.,

$$\frac{\partial V}{\partial N_{\mathrm{e}}} = \frac{\partial \mu}{\partial Q}. \tag{4}$$

As will be explained in Section 3.2, Equation (4) (and Equation 5 below) reflects the heat-engine statement of the second law of thermodynamics.

In a dilute solution, the chemical potential is $\mu = \mu_{\text{ref}} + RT \cdot \ln(cV_{\text{w}})$ [19], such that $\partial\mu/\partial Q = (RT/c)(\partial c/\partial Q) = (RT/c)(\partial c/\partial N_{\text{e}})(\partial N_{\text{e}}/\partial Q)$, where $\mu_{\text{ref}}$ is the reference chemical potential, $R$ is the gas constant, and $V_{\text{w}}$ is the molar volume of the solvent. Denote $\partial V/\partial c$ by $\delta_V$. Because $\partial V/\partial N_{\text{e}} = \delta_V(\partial c/\partial N_{\text{e}})$ and $\partial N_{\text{e}}/\partial Q = -2\xi^{-1}(\partial N_{\text{ad}}/\partial Q) = -2\Lambda/\xi$, Equation (4) can be rewritten as

$$\delta_V \triangleq \frac{\partial V}{\partial c} = -\frac{2k_{\text{B}}T}{e_0 c}\Lambda, \tag{5}$$

where $\xi$ is the Faraday constant, $\Lambda \triangleq \partial N_{\text{ad}}/\partial Q = \partial\sigma_{\text{tot}}/\partial\sigma = \partial\Gamma_{\text{ion}}/\partial\sigma$ is the charge efficiency, $\Gamma_{\text{ion}} = \sigma_{\text{tot}} - 2\sigma_0$ is the surface excess ion density, $2\sigma_0$ is the reference surface ion density of uncharged pores when $V = 0$, and $\sigma = Q/A_{\text{e}}$ is the effective surface charge density.

Charge efficiency ($\Lambda$) is an important property of the capacitive cells [20-23]. It describes how the electrolyte concentration $c$ (or the surface ion density $\sigma_{\text{tot}}$) varies with the state of charge $\sigma$. When an electrode is not charged ($V = 0$ and $\sigma = 0$), its surface adsorbs equal amounts of co-ions and counter-ions. As the electrode is charged ($V > 0$ and $\sigma > 0$), charge balance is maintained through both adsorption of additional counter-ions and repulsion of adsorbed co-ions. The former causes $\sigma_{\text{tot}}$ to increase, and the latter causes $\sigma_{\text{tot}}$ to decrease. If there were no repulsion of co-ions (i.e., if the increase in $\sigma$ were entirely associated with the adsorption of counter-ions), $\Lambda = 1$, and the increase in electrode charge would be equal to the decrease in dissolved ions in the bulk liquid phase (for $z = 1$); the predicted $\delta_V$ of the classical Gouy-Chapman-Stern (GCS) model would converge to the Nernst equation [17,20]. In general, the adsorption of counter-ions and the repulsion of co-ions occur simultaneously. As illustrated in Figure 2(b), with the swapping of co-ions and counter-ions, $\Lambda < 1$. The value of $\Lambda$ increases with the electrolyte concentration ($c$) and the surface charge density ($\sigma$), and approaches unity when $c$ and $\sigma$ are large [21-23].

### 3.2 Heat-engine statement of the second law of thermodynamics

In this section, we demonstrate that Equations (4) and (5) reflect the heat-engine statement of the second law of thermodynamics, i.e., no useful work can be produced in a cycle by absorbing heat from a single thermal reservoir [24]. Figure 3(a) depicts a capacitive-osmotic cell consisting of two porous electrodes in a dilute aqueous solution of a $z$:$z$ salt. An osmotic piston separates the solution from a reservoir of pure water. Water molecules can freely pass through the osmotic piston, while the dissolved ions are blocked. The system is closed and immersed in a thermal reservoir at $T$.

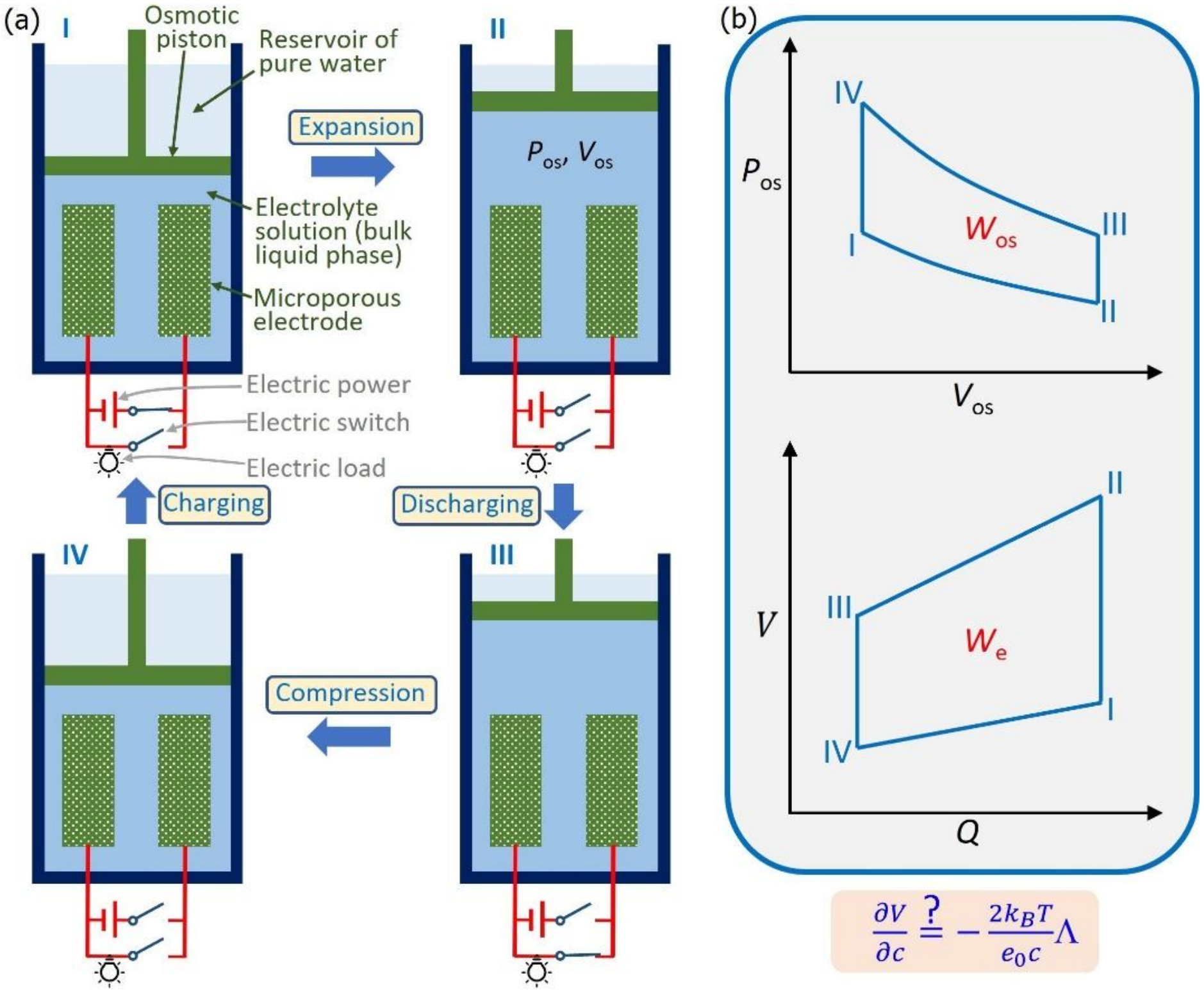

**Figure 3. (a)** A capacitive-osmotic cell. The system is closed and immersed in a thermal reservoir. The osmotic piston adjusts the electrolyte concentration in the bulk liquid phase ($c$), which in turn determines the osmotic pressure ($P_{os}$); $P_{os}$ corresponds to the in-plane pressure $P$ in Figure 1(d,e). The electric power supply controls the voltage of the porous electrodes ($V$); $V$ corresponds to the support force $F_G$ in Figure 1(d,e). **(b)** The operation cycle of the osmotic pressure ($P_{os}$) and the electric potential ($V$), analogous to Figure 1(c). Indexes I-IV indicate the system states.

Initially, at State I, the electrodes are charged with $Q$. From State I to II, both electric switches are off and the electrode charge ($Q$) remains constant; the piston moves upwards and the liquid volume in the cell ($V_{os}$) increases. As water molecules enter the cell through the osmotic piston, the electrolyte concentration ($c$) is reduced, and the cell potential ($V$) becomes higher (Figure 3b). For large-pored electrodes, the $V-c$ relationship is described by the classical GCS model: $\delta_V = -2RT\sigma/(c\xi\sqrt{\kappa_0 z^2 T^2 c + \sigma^2})$ [21-23], where $\kappa_0$ is a system parameter. For microporous electrodes, while the GCS model may not be applicable, similar phenomena have been repeatedly observed: with a constant $\sigma$, $V$ decreases as $c$ becomes larger, and vice versa [17, 25-27]. The variation in $V$ is not caused by the osmotic pressure ($P_{os}$), as $P_{os} = 2cRT$ is determined by $T$ and $c$ [28]. The increase in electrical energy ($W_{\mathrm{e}}$) and the work done by the osmotic pressure ($W_{\mathrm{os}}$) are both from heat absorbed from the environment. For $W_{\mathrm{e}}$, the heat exchange counterbalances the variation of the energy state of the adsorbed ions. This thermal-to-electric energy conversion mechanism has been extensively investigated for the capacitive concentration cells [27,29]. For $W_{\mathrm{os}}$, the system absorbs heat when the piston expands the cell, and releases heat when the cell is compressed. Such a process has been widely studied for osmotic deionization and osmotic energy conversion [30-32].

From State II to III, the lower electric switch is on, and the electrodes are discharged; the osmotic piston does not move. Since ions are released to the liquid phase, $c$ and $P_{\mathrm{os}}$ become larger. From State III to IV, both electric switches are off; the piston moves back, and the electrode charge does not change. As water is removed, $c$ and $P_{\mathrm{os}}$ increase, and $V$ decreases. Finally, from State IV to I, the upper electric switch is on. The cell is charged back to $Q$, with the liquid volume being kept constant. The system returns to its initial state. As shown in Figure 3(b), $P_{\mathrm{os}}$ consumes work $W_{\mathrm{os}}$, and $V$ produces electric energy $W_{\mathrm{e}}$. When the system is operated in the reverse direction, $W_{\mathrm{os}}$ would be produced and $W_{\mathrm{e}}$ would be consumed. According to the heat-engine statement of the second law of thermodynamics,

$$W_{\mathrm{e}} = W_{\mathrm{os}}. \tag{6}$$

Consider an isothermal cycle in which, at each step, the variation in liquid volume ($\mathrm{d}V_{\mathrm{os}}$) or the change of electrode charge ($\mathrm{d}Q$) is arbitrarily small. From State I to II, when $V_{\mathrm{os}}$ changes by $\mathrm{d}V_{\mathrm{os}}$, the work that $P_{\mathrm{os}}$ does on the environment is $P_{\mathrm{os}}\mathrm{d}V_{\mathrm{os}}$, accompanied by an equal amount of heat absorption. The cell potential increases by $-\delta_V\mathrm{d}c$, and the reduction of electrolyte concentration is $\mathrm{d}c = c\cdot\mathrm{d}V_{\mathrm{os}}/V_{\mathrm{os}}$. From State II to III, when $Q$ is reduced by $\mathrm{d}Q$, the capacitive cell does work $(V - \delta_V\mathrm{d}c)\mathrm{d}Q$ on the electric load, and the amount of the dissolved ions in the bulk liquid phase ($N_{\mathrm{e}}$) increases by $(2\Lambda/\xi)\mathrm{d}Q$. Because $c = N_{\mathrm{e}}/(2V_{\mathrm{os}})$, the electrolyte concentration increases by $\Lambda(V_{\mathrm{os}}\xi)^{-1}\mathrm{d}Q$ and $P_{\mathrm{os}}$ increases by $2RT\Lambda(V_{\mathrm{os}}\xi)^{-1}\mathrm{d}Q$. From State III to IV, $P_{\mathrm{os}}$ consumes work $-[P_{\mathrm{os}} + 2RT\Lambda(V_{\mathrm{os}}\xi)^{-1}\mathrm{d}Q]\mathrm{d}V_{\mathrm{os}}$. As the electrolyte concentration increases, the cell potential is reduced back to $V$. Finally, from State IV to I, the capacitive cell consumes work $V\cdot\mathrm{d}Q$. Overall, $W_{\mathrm{e}} = -\delta_V\mathrm{d}c\mathrm{d}Q = -(\delta_V c/V_{\mathrm{os}})\mathrm{d}V_{\mathrm{os}}\mathrm{d}Q$ and $W_{\mathrm{os}} = 2RT\Lambda(V_{\mathrm{os}}\xi)^{-1}\mathrm{d}Q\mathrm{d}V_{\mathrm{os}}$. Thus, Equation (6) can be derived from Equation (5). Inconsistency with Equation (5) is equivalent to contradiction to the heat-engine statement of the second law (Equation 6).

3.3 The second law of thermodynamics forbids intrinsic-nonequilibrium ion distribution

In this section, we confirm that, in accordance with Equation (5), the steady-state ion distribution must be in equilibrium. In other words, an intrinsic-nonequilibrium ion distribution conflicts with the heat-engine statement of the second law of thermodynamics (Equation 6).

In Figure 2(a), when the effective pore size ($d_\mathrm{e}$) is slightly larger than the effective ion size ($d_\mathrm{i}$) but less than $2d_\mathrm{i}$, the adsorbed ions in the nanopores are in quasi-1D lineups [33]. Since the charges are balanced,

$$\sigma = z(\sigma^- - \sigma^+). \tag{7}$$

The effective surface ion density in micropores ($\sigma^+$ and $\sigma^-$) may be expressed as

$$\sigma^\pm = \sigma_\mathrm{n} \exp\left(\mp \frac{ze_0V}{2k_\mathrm{B}T}\right), \tag{8}$$

where the coefficient $\sigma_\mathrm{n}$ ($\sigma_\mathrm{ct}$ or $\sigma_\mathrm{co}$) can be written as $\sigma_\mathrm{n} = \chi_0 b_\mathrm{n}(\xi\alpha_\mathrm{e}d_\mathrm{e}c)$, with subscript "n" being "ct" and "co" for counter-ion and co-ion, respectively; $\alpha_\mathrm{e}$ is the geometrical factor ($\alpha_\mathrm{e} = 0.25$ for cylindrical pores; $\alpha_\mathrm{e} = 1$ for slit pores), $\chi_0$ is the ion adsorption ratio, and $b_\mathrm{n}$ ($b_\mathrm{ct}$ or $b_\mathrm{co}$) may be a function of $V$ ($b_\mathrm{n} = 1$ when $V = 0$). If the effective concentration of confined ions in micropores ($c_\mathrm{i}$) at $V = 0$ were identical to $c$ (i.e., the micropore walls did not have any inherent capability of ion adsorption), $\sigma_\mathrm{n} = \xi\alpha_\mathrm{e}d_\mathrm{e}c$ and $\chi_0 = 1$. Generally, because the force fields are asymmetric across a solid surface, the micropore walls tend to adsorb ions even when $V = 0$ [19], such that $\chi_0 > 1$. By definition, $\chi_0$ is proportional to $D_0/D_\mathrm{i}$, where $D_0$ is the ion diffusion coefficient in the bulk liquid. The values of $b_\mathrm{n}$ depend on surface groups, surface defects, pore geometry, stericity, etc.

Equations (7) and (8) are the governing equations of the supercapacitive cell. Based on Equation (8), $\sigma^-\sigma^+ = \sigma_\mathrm{ct}\sigma_\mathrm{co}$. Substitution of it into Equation (7) gives $\sigma^\pm = (\sqrt{\sigma^2 + 4\sigma_\mathrm{ct}\sigma_\mathrm{co}z^2} \mp \sigma)/(2z)$. Hence, $\Gamma_\mathrm{ion} = \sigma^+ + \sigma^- - 2\sigma_0 = \sqrt{\sigma^2 + 4\sigma_\mathrm{ct}\sigma_\mathrm{co}z^2}/z - 2\sigma_0$. On the one hand,

$$\Lambda = \frac{\partial \Gamma_\mathrm{ion}}{\partial \sigma} = \frac{\sigma + 2z^2\delta_\sigma}{z\sqrt{\sigma^2 + 4\sigma_\mathrm{ct}\sigma_\mathrm{co}z^2}}, \tag{9}$$

where $\delta_\sigma \triangleq \partial(\sigma_\mathrm{ct}\sigma_\mathrm{co})/\partial\sigma$. On the other hand, based on the $V - \sigma$ relationship determined by Equations (7) and (8), with $\sigma_\mathrm{co} = \chi_0 b_\mathrm{co}(\xi\alpha_\mathrm{e}d_\mathrm{e}c)$,

$$\delta_V = \frac{\partial V}{\partial c} = -\frac{2k_\mathrm{B}T}{ze_0c}\frac{\sigma}{\sqrt{\sigma^2 + 4\sigma_\mathrm{ct}\sigma_\mathrm{co}z^2}}. \tag{10}$$

At low electrolyte concentrations ($c < 20$ mM), the adsorbed ions tend to diffuse individually in the nanopores. Therefore, compared with the order-of-magnitude variations in effective surface ion density caused by $V$ [10-16], the influence of $c$ on $b_\mathrm{n}$ is neglected in the derivation of Equation (10). This simplification does not affect our experiment in Sections 4 in determining whether the system is in equilibrium or nonequilibrium. A comparison of Equations (9) and (10) shows that, to satisfy Equation (5), we must have

$$\delta_\sigma = 0. \tag{11}$$

Equation (11) requires $\sigma_\mathrm{ct}\sigma_\mathrm{co}$ and $\sigma$ to be uncorrelated, i.e., $\bar{\sigma}_\mathrm{ct}\bar{\sigma}_\mathrm{co} = 1$, which corresponds to $D_\mathrm{i}$ being independent of $V$. Under this condition, $\sigma^+$ and $\sigma^-$ are proportional to the Boltzmann factor.

When $\delta_\sigma = 0$, Equation (8) represents the equilibrium ion distribution, which reflects the entropy statement of the second law of thermodynamics (i.e., the principle of maximum entropy). As discussed in Section 3.2, Equation (5) represents the heat-engine statement of the second law of thermodynamics. In essence, Equation (11) requires that Equation (8) and Equation (5) are equivalent, as they should be.

However, the computer simulations in [10-16] suggest that $D_\mathrm{i}$ significantly changes with $V$ and therefore $\delta_\sigma \neq 0$, contradicting Equation (11). That is, the steady-state ion distribution is spontaneously out of equilibrium. For the capacitive-osmotic cell in Figure 3, similarly to the models in Figure 1, the intrinsic-nonequilibrium steady state cannot be explained in the conventional framework of statistical mechanics. If $\delta_\sigma \neq 0$, Equations (9) and (10) indicate that $|\delta_V| \neq 2k_\mathrm{B}T\Lambda/(e_0 c)$, such that $W_\mathrm{e} \neq W_\mathrm{os}$ in the isothermal cycle in Figure 3.

## 4. Experimental procedure and results

Our experiment focuses on Equations (5) and (11); heat exchange and work production are not directly measured. The experiment is performed on a microporous carbon in aqueous cesium pivalate (CsPiv) solutions (Figure 4a). The pore size of the carbon ($d_\mathrm{e}$) is ~1 nm [34]. The effective pivalate ion size ($d_\mathrm{i}$) is ~0.7 nm (Figure 4b). The experimental procedure is similar to that used in the previous study of the same carbon in sodium chloride (NaCl) solutions [17].

The capacitive cell in Figure 4(a) was slowly charged at ambient temperature (~22 °C) with a constant current ($I$). The initial CsPiv concentration ($c$) was 10 mM, 12 mM, 14 mM, or 16 mM. For each initial $c$, as the electrode charge ($Q$) was raised, the increase in cell potential ($V$) (Figure 5a) and the decrease in $c$ (Figure 5b) were recorded; $c$ was determined through the liquid conductivity ($\eta$), monitored by the nickel-wire probe embedded in the cell. The rate convergence test (Figure 4c) suggests that the testing data reflect the steady state (see Section 6.1 for more discussion). The charge efficiency ($\Lambda$) is calculated from the measured $c$ and $Q$ (Figure 5c). By comparing the $V - Q$ curves of different initial $c$ in Figure 5(a), we obtain the concentration sensitivity of electric potential, $|\delta_V|$, as shown by the hollow symbols in Figure 5(d). The details of the experimental setup and procedure are given in Sections 4.1-4.6 below.

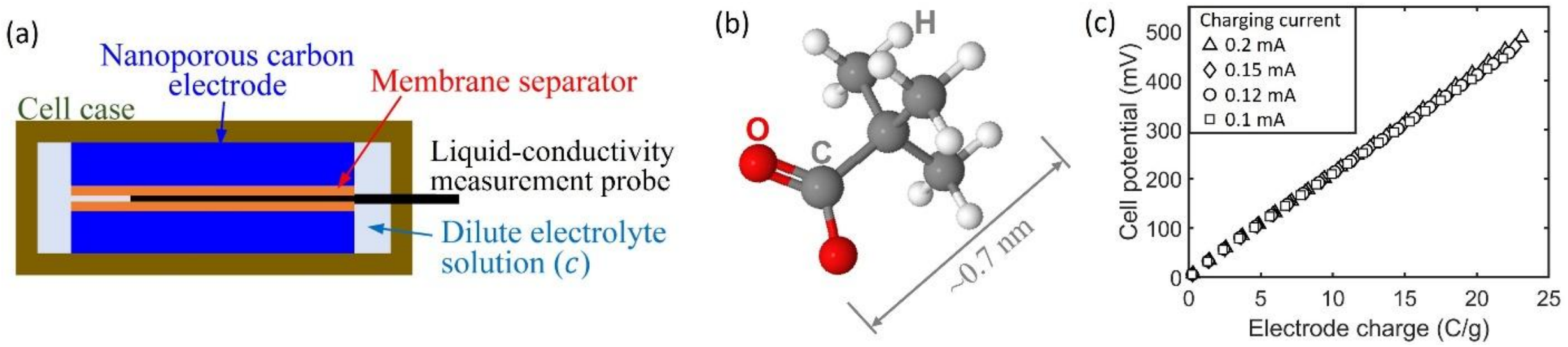

**Figure 4.** Schematics of **(a)** the supercapacitive cell and **(b)** a pivalate ion. **(c)** The rate convergence test result, suggesting that the measurement data reflect the steady state (see Section 6.1 for more discussion).

4.1 Preparation of materials and cell components

The electrodes were made of Spectracarb-2225 Type-900 carbon, a microporous material commonly used in studies of double-layer supercapacitors [e.g., 17,34]. The as-received carbon cloth was cut into electrode discs with a diameter of 15.88 mm, and dried in a VWR-1330GM oven for 24 h at 120 °C. The electrode mass ($m_e$) was around 25 mg. Two identical electrode discs were soaked in 20 mL CsPiv solution in a VWR Shel-Lab-1410 vacuum oven for 5 min at 94.8 kPa. Dreamweaver Titanium-30 porous membrane separators were cut into discs and immersed in the same aqueous solution of CsPiv for 10 min. The membrane disc diameter was 17.46 mm. From a 415-μm-thick McMaster-85585K15 polycarbonate film, two spacer rings were cut using a punch head, with inner and outer diameters of 7.1 and 15.9 mm, respectively.

The cell case was made of two circular McMaster-1221T63 polyacrylic discs, with a thickness of 25.4 mm and a diameter of 76.2 mm (Figure 6). One disc was used as the top case and the other was the bottom case. On each cell case, eight edge holes were drilled by using a McMaster-28015A51 Palmgren drill press with a 7.1-mm-diameter drill bit (McMaster 2901A126). The holes were equally spaced, with the center-to-center distance being 50.8 mm along the diagonal direction. A liquid-replacement hole was drilled at the center of each case using a 3.18-mm-diameter drill bit (McMaster 2901A115).

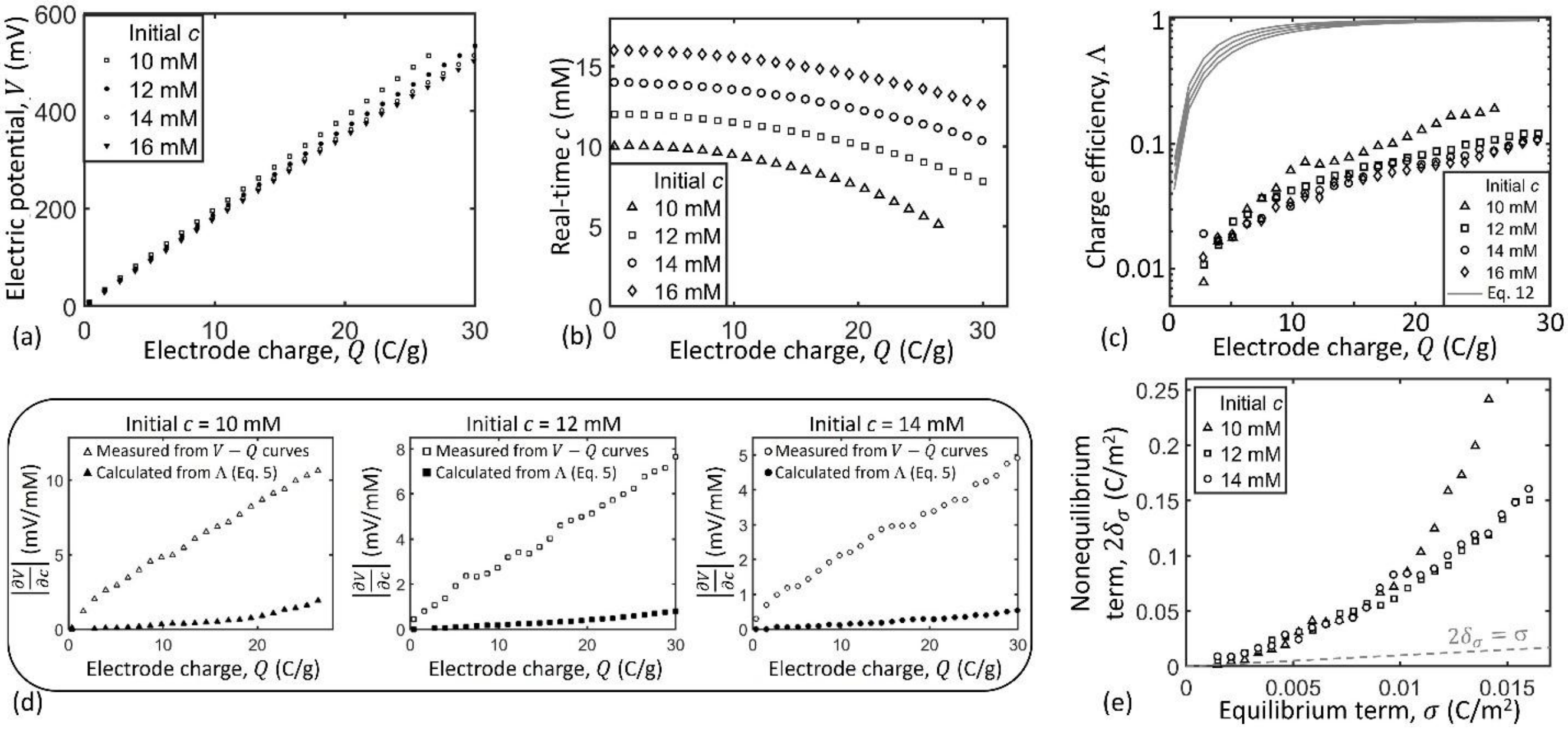

**Figure 5. (a)** The cell potential ($V$) increases as the cell is charged at a constant current. The charge curves are influenced by the initial CsPiv concentration ($c$). **(b)** The measured real-time CsPiv concentration ($c$), as a function of the electrode charge ($Q$). When the cell is charged, $c$ keeps decreasing. **(c)** The charge efficiency (Λ). As the cell is charged, Λ keeps increasing. The hollow symbols are directly obtained from the $c - Q$ data in panel (b); the solid curves are calculated from Equation (12). They differ by one order of magnitude. The four solid curves correspond to initial $c$ of 10, 12, 14, and 16 mM, respectively; a smaller initial $c$ gives a higher curve. **(d)** Comparison of the $|\delta_V|$ values directly measured from the $V - Q$ data in panel (a) (the hollow symbols) and computed from the Λ data in panel (c) (the solid symbols). They differ by one order of magnitude. The initial $c$ is 10 mM (left), 12 mM (middle), or 14 mM (right). **(e)** The intrinsic-nonequilibrium term in the numerator of Equation (9), $2\delta_\sigma$ (for $z = 1$). The horizontal axis is the equilibrium term in the numerator of Equation (9), i.e., the effective surface charge density $\sigma = Q/A_e$.

Each connection tube was fabricated by inserting a 200-mm-long ethylene-vinyl acetate (EVA) tube (McMaster 1883T1) into a 50-mm-long McMaster-5231K124 poly(vinyl chloride) (PVC) tube. The interface between the inner and outer tubes was sealed with McMaster-7605A18 J-B Weld epoxy and cured for 10 h at ambient temperature. Two connection tubes were attached to the central holes of the top and bottom cell cases, respectively. A 0.6-mm-diameter BD PrecisionGlide 305194 needle was tightly inserted into the EVA tubing, and the other end of the needle was connected to a 1-mL BD 309659 syringe.

A Panasonic EYG-S121803 graphite sheet was cut into 1.5-mm-wide 20-mm-long strips using a McMaster-3962A48 steel razor blade. A 25-µm-thick MTI MF-NiFoil-25u nickel foil was cut into 30-mm-long 2-mm-wide strips and repeatedly flattened using a Durston DRM F150 rolling mill with a roller gap of 20 μm. A graphite strip was affixed to a nickel strip with 4-mm-wide McMaster-7648A32 Kapton tape to form an electrical outlet tab. One tab was attached to the bottom cell case using the same Kapton tape, with an overlap length of 10 mm. Another tab was similarly attached to the top cell case.

Two 50-mm-long 0.5-mm-wide nickel strips were cut from the nickel foil and repeatedly flattened using the rolling mill, with a roller gap of 20 μm. A liquid-conductivity measurement probe was fabricated by placing the two nickel strips in parallel. The gap between the strips was 1.0 mm, and the gauge length was 10 mm. The strips were secured with three 0.8-mm-wide Kapton tapes and sandwiched between two membrane separators.

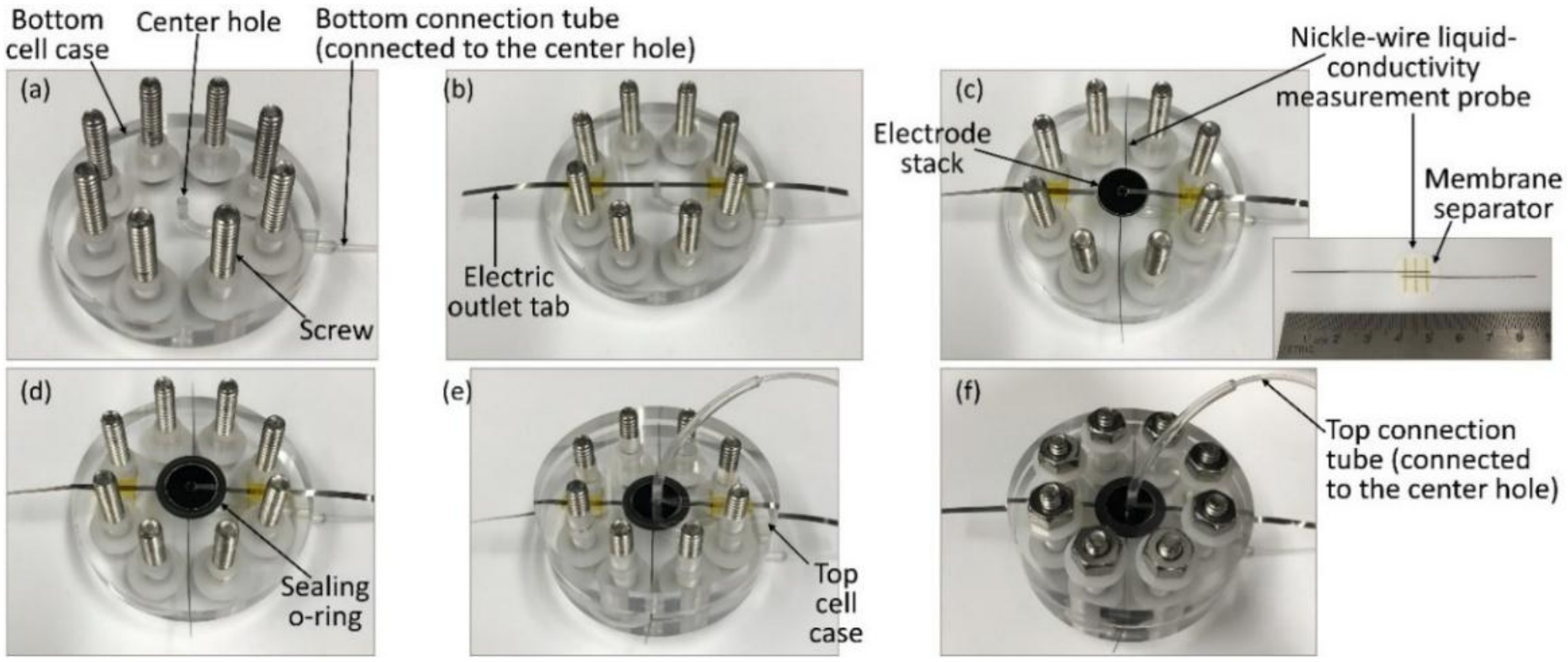

**Figure 6.** The cell assembly process: **(a)** the bottom cell case; **(b)** the electric outlet tabs; **(c)** the electrodes and the nickel-wire liquid-conductivity measurement probe; the inset at the lower-right corner shows a close-up view of the nickel-wire probe; **(d)** the sealing o-ring; **(e)** the top cell case; **(f)** the assembled supercapacitive cell. The diameter of the cell case is 76.2 mm.

4.2 Cell formation

As shown in Figure 6, eight McMaster-92196A821 stainless steel screws were inserted in the edge holes of the bottom cell case, together with McMaster-95606A420 nylon washers. One carbon electrode was placed on the bottom case at the center, and covered by the probe-embedded membrane separators, the other carbon-disc electrode, and the top cell case. The electric connections were optimized by adjusting the tab position. In between the electrode stack and the top cell case, there were two layers of spacer rings. On the screws at the bottom case, two layers of McMaster-90295A450 nylon shims were added. Surrounding the electrode stack, there was a McMaster-8297T174 Viton o-ring. McMaster-91849A029 stainless-steel nuts were fastened onto the screws, together with McMaster-95606A420 nylon washers. The fastening was performed by a McMaster-5503A22 L-key and a McMaster-7152A812 wrench. In the assembled cell, the shims contacted both the top case and the bottom case. Excess electrolyte was extracted by two syringes through the connection tubes.

The electric outlet tabs were connected to a Neware CT-ZWJ-4S-T Analyzer. The cells were cycled from 0 to 800 mV at 0.1 mA for 20 cycles. If the coulombic efficiency of the cell was lower than 98% or the cell impedance was higher than 60 Ω, the cell would be rejected. Coulombic efficiency is defined as the ratio of the discharge capacity to the charge capacity.

4.3 Rate convergence test

The capacitive cell was tested in charge-discharge cycles. For different cycles, the current ($I$) was 0.2 mA, 0.15 mA, 0.12 mA, or 0.1 mA. A Hanna HI5321-01 Meter was connected to the nickel-wire liquid-conductivity measurement probe, so that the liquid conductivity ($\eta$) was measured simultaneously with $V$. The initial CsPiv concentration was 10 mM.

In each charge-discharge cycle, the current was periodically stopped. Each stop lasted 1 min, after which the charging current was resumed. The maximum $Q$ was 0.48 C. At each charging rate, the cell was first pre-cycled for five cycles. Then, the voltage was maintained at 0 V for 2 min. During the sixth cycle, the cell potential and liquid conductivity were recorded after the resting period at each stop.

Figure 4(c) shows the measured $V-Q$ profiles. When the current $I=0.1$ mA, it took ~80 min to complete the charging process. When $I$ was 0.12 mA, 0.15 mA, or 0.2 mA, the charging duration ($t_c$) was ~67 min, ~53 min, or ~40 min, respectively. The results suggest that, at the slow charging rates, the process was close to steady state. Further discussion is provided in Section 6.1.

4.4 Charge-discharge cycle

A cell was prepared with an initial $c$ of 10 mM, and charged at ambient temperature (~22 $^{o}$C) with $I=0.1$ mA. The current was periodically stopped. At each stop, the system was allowed to rest for 1 min, followed by measurement of the cell potential ($V$) and the liquid conductivity ($\eta$); the current was then resumed.

After each charge-discharge cycle, the liquid was replaced (see Section 4.5 below) and the initial $c$ was increased by 2 mM. The measurements of $V$ and $\eta$ were then repeated with the new electrolyte solution. In total, the liquid phase was replaced three times. That is, the CsPiv concentration at the onset of cell charging was 10, 12, 14, or 16 mM (see Figure 5b).

## 4.5 Liquid replacement

In Figure 3(a), the electrolyte concentration ($c$) is adjusted by a frictionless osmotic piston. Experimentally implementing this mechanism is difficult because of piston friction and the imperfect ion separation of available semi-permeable membranes. To circumvent these hurdles, liquid replacement has been commonly employed in studies of capacitive desalination [22-25] and concentration cells [27,29] to precisely control the electrolyte concentration. This process has little effect on other cell components, such as the electrode layers and the electrical contacts. It is performed by slowly flowing a new electrolyte solution through the cell, often across the electrode stack. Once the steady state is established, the original liquid is replaced by the new solution [17].

The center hole of the top CsPiv cell case was connected to a 60-mL syringe, which contained the new CsPiv solution. The CsPiv concentration of the new liquid was 2 mM higher than that of the original solution in the cell. An Instron-5582 machine pressed the syringe at a constant rate, and a slow flow was injected into the cell for about 30 min. The flow rate was approximately 4 mL/min. The excess liquid flowed out of the cell from the center hole in the bottom cell case. The liquid conductivity ($\eta$) was measured continuously by the embedded nickel-wire probe. Once $\eta$ was stabilized at the new level, the flow was stopped.

After resting for ~30 min, the cell was pre-cycled for five cycles between 0 and 800 mV. Testing on the cell would be discontinued if the coulombic efficiency was below 98% or the internal impedance exceeded 60 Ω.

## 4.6 Measurement results

Figure 5(a) shows the increase in cell potential ($V$) when the cell was charged, where $Q = 3.6It_c/m_e$, $m_e$ is the mass of electrode in gram, and $t_c$ is the charging duration in hr. Figure 5(b) shows the CsPiv concentration ($c$) profiles during charging, obtained from the measured liquid conductivity ($\eta$) as $c = 0.1664\eta^2 + 1.2816\eta - 0.0934$, with a coefficient of determination of 0.9999. The units of $c$ and $\eta$ are mM and mS/cm, respectively. The $c - \eta$ relation was calibrated by measuring $\eta$ for various known $c$; $c$ was adjusted through liquid replacement from 1 mM to 50 mM (see Section 4.5), with an interval of 2 mM when $c \leq 20$ mM and 10 mM when $c > 20$ mM.

Figure 5(c) shows the charge efficiency, calculated from Figure 5(b) as $\Lambda = -\xi V_{\rm L}\Delta c/\Delta Q$ [35], where $V_{\rm L} = V_{\rm cell} - V_{\rm C} - V_{\rm SP}$ is the liquid volume, $V_{\rm cell} = A_{\rm c}d_{\rm c}$ is the cell volume, $A_{\rm c}$ = 197.9 mm$^2$ is the cross-sectional area of the electrode, $d_{\rm c}$ is the height of the cell cavity, $V_{\rm C} = m_{\rm e}/\rho_{\rm C}$ is the volume of the solid carbon, $V_{\rm SP} = m_{\rm SP}/\rho_{\rm SP}$ is the solid volume of the membrane separators, $\rho_{\rm C} = 2.2$ g/cm$^3$ is the density of carbon, $m_{\rm SP}$ is the mass of the membrane separators, $\rho_{\rm SP}$ is the density of the membrane; $\Delta c$ is the change in $c$ between two consecutive measurements of $\eta$ along the $c - Q$ curve, and $\Delta Q$ is the corresponding increase in electrode charge.

Figure 5(d) shows the ion-concentration sensitivity of cell potential, $|\delta_V|$. The hollow symbols are obtained directly from Figure 5(a). Between two adjacent $V - Q$ curves, at the same $Q$, the difference in $V$ is divided by the difference of real-time $c$ (Figure 5b). The solid symbols are computed from Equation (5) as $|2k_{\rm B}T\Lambda/(e_0c)|$, using the values of $\Lambda$ in Figure 5(c). To be conservative, $c$ is taken as the lower bound of the range of CsPiv concentration involved in the calculation.

In Figure 5(e), based on Equations (9) and (10), the intrinsic-nonequilibrium term in the numerator of Equation (9) is computed as $2\delta_\sigma = [-2k_{\rm B}T\Lambda/(e_0c\delta_V) - 1]\sigma$ (for $z = 1$), where $\Lambda$ is from Figure 5(c), $\delta_V$ is from Figure 5(d), the equilibrium term in the numerator of Equation (9) is $\sigma = Q/A_{\rm e}$, and $A_{\rm e} = 1870$ m$^2$/g is the Brunauer-Emmett-Teller (BET) specific surface area [34]. The BET value of $A_{\rm e}$ represents the upper limit of the specific surface area of the carbon and is used here for an order-of-magnitude estimate of $\sigma$. The real-time $c$ is obtained from Figure 5(b).

## 5. Analysis of the experimental data

### 5.1 Incompatibility with the second law of thermodynamics

Remarkably, the measured $\Lambda$ and $|\delta_V|$ do not satisfy Equation (5), as demonstrated in Figure 5(d). The solid symbols are $2k_{\rm B}T\Lambda/(e_0c)$, calculated from the testing data of $\Lambda$ in Figure 5(c); they are the right-hand side of Equation (5). The hollow symbols are $|\delta_V|$, directly obtained from Figure 5(a); they are the left-hand side of Equation (5). The three panels are for different initial CsPiv concentrations ($c$): from left to right, the initial $c$ are 10 mM, 12 mM, and 14 mM, respectively. In all three cases, $|\delta_V|$ is greater than $2k_\mathrm{B}T\Lambda/(e_0c)$ by nearly one order of magnitude. Since Equation (5) no longer holds, if the supercapacitive cell is operated in the isothermal cycle in Figure 3, the net produced work would be greater than the net consumed work. Such an unusual effect is incompatible with the heat-engine statement of the second law (Equation 6).

The reference test on NaCl solutions [17] has confirmed that the testing setup and procedure are adequate for the study on Equation (5) (see Section A3 of the Appendix). With the small $Na^+$ and $Cl^-$ ions, the system behavior can be well described by the second law. Yet, for CsPiv, as the effective ion size ($d_\mathrm{i}$) is less than the effective micropore size ($d_\mathrm{e}$) but larger than $d_\mathrm{e}/2$, conventional theory is inapplicable (see Section 6.2).

In fact, Figure 5(d) is not the first experimental evidence of non-thermodynamic phenomena of macroscopic systems. In a study on nonwetting liquids in nanoporous particles [36,37], large temperature changes were observed [4]. Across a nanoporous membrane one-sidedly surface-treated with bendable organic chains, an asymmetric crossing ratio was measured [5]. The previous experiments were based on time-dependent locally nonchaotic entropy barriers. The present investigation focuses on the energy-barrier SND.

## 5.2 Intrinsic-nonequilibrium steady-state ion distribution

In an equilibrium system, the effective surface ion density is proportional to the Boltzmann factor (Equation 8). With $\delta_\sigma = 0$, $\bar{\sigma}_\mathrm{ct}\bar{\sigma}_\mathrm{co} = 1$ and Equation (9) is reduced to (for $z = 1$)

$$\Lambda = \frac{Q}{\sqrt{Q^2+4(\varphi\xi d_\mathrm{e}c)^2}}, \tag{12}$$

where $\varphi = \chi_0\alpha_\mathrm{e}A_\mathrm{e}$. The ion adsorption ratio $\chi_0$ can be estimated as $c_\mathrm{i}/c$ at $V = 0$. In the reference experiment on NaCl solutions [17], when $\chi_0\alpha_\mathrm{e} = 2.9$, the measured $\Lambda$ was in agreement with Equation (12) (see Section A3 of the Appendix). This value of $\chi_0\alpha_\mathrm{e}$ is consistent with the literature data of carbon adsorbents [38].

To assess $\chi_0$ for CsPiv, we performed an ion adsorption experiment. A 0.25-g pristine carbon disk was immersed in 20 mL 10-mM CsPiv solution. After 1 h, $c$ converged to 9.90 mM. According to [34], in the carbon electrode, the specific volume of the nanopores larger than the ion size (~7 Å) is about 0.16 cm$^3$/g. Since $\alpha_e = 0.25$ for cylindrical pores, $\chi_0\alpha_e = (c_i/c)\alpha_e \approx 1.25$. To be conservative, we take $A_e$ as the BET specific area (the upper limit of the specific surface area), 1870 m$^2$/g [34]. Thus, $\varphi = \chi_0\alpha_e A_e \approx 2.3 \times 10^3$ m$^2$/g.

As shown in Figure 5(c), with such a $\varphi$, Equation (12) fails to describe the experimental measurement. The predicted solid lines are different from the hollow symbols by one order of magnitude. To keep the calculated Λ on the same scale as the testing data, $\varphi$ must be greatly increased to 1.1×10$^5$ m$^2$/g, which corresponds to an unacceptably large $\chi_0$ around $240$, much exceeding the ion adsorption capacity of nanoporous carbons by two orders of magnitude [34]. It is all the more unsatisfactory that, even if we set $\chi_0 = 240$, the computed curves merely reproduce the qualitative trend of the $\Lambda - Q$ relationship (i.e., Λ increases with $Q$). Since in an electrolyte solution the effective specific surface area of nanoporous carbon is significantly smaller than the BET value, the actual required $\chi_0$ tends to be much larger than 240.

It is clear that, to analyze Λ, Equation (12) should not be used. That is, the ion distribution is far from equilibrium, and $\delta_\sigma \neq 0$ must not be ignored in Equation (9). In Figure 5(e), when the electrode charge is small, the intrinsic-nonequilibrium term in the numerator of Equation (9) ($2z^2\delta_\sigma$) is comparable with the equilibrium term ($\sigma$). The degree of intrinsic nonequilibrium drastically increases as the electrodes are charged. When $\sigma$ reaches about 15 mC/m$^2$, $2z^2\delta_\sigma$ is more than 10 times greater than $\sigma$. The large $2z^2\delta_\sigma$ is compatible with the MD simulation results [10-12] that as $V$ rises, $D_i$ can increase by 1-2 orders of magnitude. The nonequilibrium characteristics are intrinsic. The experimental data in Figure 5 reflect the steady state (see Section 6.1 for the details). Under isothermal conditions and with a given electrode charge state ($\sigma$), the non-Boltzmann ion distribution is not sustained by any external thermodynamic driving force. In other words, inherently, the system cannot relax to equilibrium (see Section 6.2 for further discussion). The large discrepancy between the measured Λ and the prediction of equilibrium thermodynamics (Equation 12) is the direct cause of the imbalance in the experimental data of $\delta_V$ in Figure 5(d).

## 6. Discussion

### 6.1 The measurement result reflects the steady state

Figure 5 reflects the steady state of the CsPiv cells. First, the rate-convergence test (Figure 4c) indicates that the measurement data are insensitive to the charging rate. This result rules out kinetic interpretations of the measured intrinsic-nonequilibrium ion distribution, such as potential-dependent electrosorption barriers or partial conformational relaxation of confined ions.

Second, the reference NaCl experiment [17] demonstrated that, for small ions, the testing results can be described by conventional thermodynamic analysis (see Section A3 of the Appendix). For CsPiv, the low charge efficiency is associated with a mild ion-concentration gradient, such that the liquid-conductivity measurement should be more accurate than that of NaCl. Moreover, because $D_{\mathrm{i}}$ of large ions is greater than that of small ions [10-12], the CsPiv cells tend to reach the steady state more rapidly than the reference NaCl cells.

Third, with the slow charging rate, the typical charging duration ($t_{\mathrm{c}}$) is ~80 min. The ion diffusivity of CsPiv can be estimated as $D_0 = RTc/(\xi^2 z^2 \eta)$ [39], which is ~1200 μm$^2$/sec for $c$ = 10 mM. With the characteristic length being the electrode half-thickness $L_0 \approx 200$ μm, the time constant of ion diffusion is $\tau_0 = L_0^2/D_0 \approx 33$ sec. It is shorter than $t_{\mathrm{c}}$ by over 2 orders of magnitude. This assessment is conservative, since $D_{\mathrm{i}}$ in micropores is much larger than $D_0$ [10-16] and therefore the actual $\tau_0$ is much less than 33 sec.

Fourth, if the transient effect were not negligible, co-ion repulsion would be incomplete and, to maintain charge balance at the electrodes, more counter-ions would have to be adsorbed. Consequently, the charge efficiency and the predicted $|\delta_V|$ by Equation (5) would be even smaller than the measured values. Fifth, the coulombic efficiency (> 98%) achieved in the experiment is high (see Section 4.2). Finally, compared with the bulk liquid phase, ion motion is much faster in the strong Coulomb field near the electrode surfaces [40,41].

### 6.2 Mechanisms of the intrinsic-nonequilibrium ion distribution

The experiment on the CsPiv cells (Figure 5c-e) and the MD simulations of $D_{\mathrm{i}}$ [10-16] indicate the presence of strong intrinsic-nonequilibrium mechanisms in the microporous electrodes; that is, the micropores act as the SND. In Equation (9), the intrinsic-nonequilibrium term ($2z^2\delta_\sigma$) tends to be an order of magnitude more important than the equilibrium term $\sigma$ (Figure 5e), which leads to the large mismatch between the test data of $\Lambda$ and the equilibrium prediction (Figure 5c). In Equation (8), dividing the surface ion density ($\sigma^\pm$) by the Boltzmann factor ($e^{\mp\beta z e_0 V/2}$) yields a coefficient $\sigma_\mathrm{n}$ that highly depends on $V$.

The numerical studies in [10,11] suggest that the intrinsic-nonequilibrium surface ion density originates from confinement effects of the nanopore walls, including accelerated diffusion and constrained relaxation of ion configurations. Other factors may also contribute, such as surface heterogeneities, ion-water interactions, differences in diffusion rates [12], etc. For example, it is well known that carbon surfaces have a large number of defects [42] and charged groups [43], which can dominate the properties of carbon. These surface heterogeneities cause local variations in potential and energy exchange, denoted by $E_\mathrm{t}$. Because the effective nanopore size $d_\mathrm{e}$ is only slightly larger than the effective ion size $d_\mathrm{i}$, when the adsorbed ions diffuse along the axial direction in a nanopore, $E_\mathrm{t}$ cannot be circumvented.

In an equilibrium three-dimensional (3D) system, the effective local surface ion density corresponding to $E_\mathrm{t}$ is $\sigma_\mathrm{t} = \alpha_\mathrm{h}(d_\mathrm{e} c \delta_0)\delta_\mathrm{t}$, where $\delta_\mathrm{t} = e^{-\beta z e_0 E_\mathrm{t}}$ is the Boltzmann factor, and $\alpha_\mathrm{h}$ is the coefficient related to the pore geometry, the steric effect, etc. Yet, under the condition of $d_\mathrm{i} < d_\mathrm{e} < 2d_\mathrm{i}$, the motion of the adsorbed ions is quasi-1D in the nanopores. Since only the axial-dimension thermal movement ($v_\mathrm{a}$) is directly relevant to the interaction with $E_\mathrm{t}$, the ion transmission ratio is governed not by $\delta_\mathrm{t}$, but rather by $\delta_2 = \int_{v_\mathrm{cr}}^{\infty} p_\mathrm{a}(v_\mathrm{a})\mathrm{d}v_\mathrm{a} = 1 - \mathrm{erf}(\sqrt{\beta z e_0 E_\mathrm{t}})$, where $v_\mathrm{cr} = \sqrt{2 z e_0 E_\mathrm{t}/m_\mathrm{i}}$, $p_\mathrm{a}(v_\mathrm{a}) = \sqrt{\beta m_\mathrm{i}/(2\pi)}\, e^{-\beta m_\mathrm{i} v_\mathrm{a}^2/2}$ is the 1D Maxwell-Boltzmann distribution of $v_\mathrm{a}$, and $m_\mathrm{i}$ is the ion mass. Hence, $\sigma_\mathrm{t} \propto (d_\mathrm{e} c \delta_0)\delta_2$. Such a confinement effect of nanopore walls is comparable to the role of the nonchaotic stage walls in Figure 1(d,e), and $\delta_2$ is the counterpart of $\delta_1$.

The intrinsic-nonequilibrium $\sigma_\mathrm{t}$ can be further analyzed using the following two thermodynamic forces: $F_\mathrm{t} = z e_0 A_\mathrm{t} \sigma_\mathrm{t}$ and $P_\mathrm{t} = k_\mathrm{B} T N_\mathrm{r}/V_\mathrm{os}$, where $A_\mathrm{t}$ and $N_\mathrm{r}$ are the effective area and the ion number associated with $E_\mathrm{t}$, respectively. The conjugate variables of $F_\mathrm{t}$ and $P_\mathrm{t}$ are $E_\mathrm{t}$ and $-V_\mathrm{os}$, respectively. Compared with Figure 1(d,e), the form of $F_\mathrm{t} = z e_0 A_\mathrm{t} \sigma_\mathrm{t}$ is similar to $F_\mathrm{T} \propto mgN_\mathrm{T}$, and the form of $P_\mathrm{t} = k_\mathrm{B} T N_\mathrm{r}/V_\mathrm{os}$ is similar to $P = k_\mathrm{B} T N_\mathrm{G}/A_\mathrm{G}$; $E_\mathrm{t}$ is the counterpart of $mgz_\mathrm{t}$, and $V_\mathrm{os}$ is the counterpart of $A_\mathrm{P}$. Equation (3) requires that $-\partial F_\mathrm{t}/\partial V_\mathrm{os} = \partial P_\mathrm{t}/\partial E_\mathrm{t}$. If $E_\mathrm{t}$ slightly varies by $\mathrm{d}E_\mathrm{t}$, the variation in adsorbed ions is $A_\mathrm{t}(\partial\sigma_\mathrm{t}/\partial E_\mathrm{t})\mathrm{d}E_\mathrm{t}$ and accordingly, $N_\mathrm{r}$ changes by the same amount; thus, $\partial N_\mathrm{r}/\partial E_\mathrm{t} = -A_\mathrm{t}(\partial\sigma_\mathrm{t}/\partial E_\mathrm{t})$. Since $c = N_\mathrm{e}/(2V_\mathrm{os})$, when $\sigma_\mathrm{t} = \alpha_\mathrm{h}(d_\mathrm{e}c\delta_0)\delta_\mathrm{t}$, $-\partial F_\mathrm{t}/\partial V_\mathrm{os} = ze_0A_\mathrm{t}\sigma_\mathrm{t}/V_\mathrm{os}$, which is equal to $\partial P_\mathrm{t}/\partial E_\mathrm{t} = -k_\mathrm{B}TA_\mathrm{t}V_\mathrm{os}^{-1}(\partial\sigma_\mathrm{t}/\partial E_\mathrm{t})$. Thus, the solution of Equation (3) is the equilibrium ion distribution ($\sigma_\mathrm{t} \propto \delta_\mathrm{t}$). Equivalently speaking, the intrinsic-nonequilibrium ion distribution cannot satisfy Equation (3).

Another intrinsic-nonequilibrium mechanism may involve the excess ion energy, $K_\mathrm{t}$. In the quasi-1D nanoenvironment, the $K_\mathrm{t}$-related ion diffusion ratio can be assessed as $\overline{D}_\mathrm{n} = \int_0^{v_\mathrm{t}} p_\mathrm{a}(v_\mathrm{a})\mathrm{d}v_\mathrm{a} = \mathrm{erf}(\sqrt{\beta K_\mathrm{t}})$, where $v_\mathrm{t} = \sqrt{2K_\mathrm{t}/m_\mathrm{i}}$. In comparison, its equilibrium counterpart is $\overline{D}_\mathrm{e} = 1 - e^{-\beta K_\mathrm{t}}$. Since $\beta K_\mathrm{t} > 0$, $\overline{D}_\mathrm{n} > \overline{D}_\mathrm{e}$, consistent with the MD simulation results that $D_\mathrm{i}$ increases drastically with $V$ [10-16]. With a relatively small $\beta K_\mathrm{t}$, $\overline{D}_\mathrm{n}$ may be larger than $\overline{D}_\mathrm{e}$ by more than one order of magnitude.

In essence, the confined ions form the locally nonchaotic SND, like the Knudsen-gas particles in the low-height steps in Figure 1. SND cannot reach thermodynamic equilibrium, resulting in the overall non-Boltzmann steady state. Because the micropore walls separate the interior of the electrode from the bulk liquid phase, energy and mass can only be transported along the axial direction, and the adsorbed ions individually interact with the inner electrode surfaces.

The intrinsic-nonequilibrium state demonstrated in Figure 5 should not be attributed to electrical energy dissipation. First, as the non-Boltzmann ion distribution is measured at steady state with fixed $V$ and $\sigma$, no matter how the steady state is reached, the conventional boundaries of the second law have already been broken. The Coulomb force is the counterpart of the gravitational force in the "toy model" in Figure 1(d,e), and the charged nanopores are analogous to the low-height stages. Second, even if certain ion reorientation or reconfiguration processes might dissipate some electrical energy (i.e., are not perfectly "friction-free"), the associated energy loss must be less than that of the hysteresis of the charge-discharge curves (< 2%). On the one hand, the supercapacitive cells exhibit high coulombic efficiency (> 98%) and good linearity. On the other hand, the differences in cell potential of the charge curves in Figure 5(a) are unusually large: between initial $c$ values of 10 and 12 mM and of 12 and 14 mM, the variations in $V$ are about 6.8% and 3.5%, respectively; both are much greater than 2% and neither can be explained by hysteresis. Third, in the MD simulations in [10-16], the large effective diffusion coefficient of confined ions ($D_\mathrm{i}$) is observed without dissipative mechanisms, under steady-state conditions at a given state of charge. The processes associated with $D_\mathrm{i}$ are thermally driven. Fourth, a larger $D_\mathrm{i}$ leads to a decrease in the effective surface ion density in the nanopores, such that the ion concentration in the bulk liquid phase increases; consequently, at the same state of charge, the cell potential is lower. That is, a large $D_{\mathrm{i}}$ tends to reduce $V$ rather than consume additional electrical energy. Moreover, the testing setup and procedure have been validated by the reference experiment on the NaCl cells using the same carbon [17]. Regardless of the specific behaviors of the adsorbed ions, their effects on charge curves are reflected in Figure 5(a) and are therefore accounted for in the data analyses of Figure 5(c-e). The variation in electrical energy is balanced by the heat exchange with the environment (the thermal reservoir).

## 6.3 Fundamental difference from the Carnot cycle and the conventional nonequilibrium phenomena

At first glance, the $V-Q$ cycle in Figure 3(b) is somewhat similar to the operation of a Carnot engine. From state II to III, the cell is discharged and does work on the environment. From state IV to I, the cell is charged and consumes work. Because the discharging voltage is higher than the charging voltage, the net work produced is positive.

In a Carnot engine, the difference between charging and discharging curves is achieved by varying temperature, and the system performance is limited by the Carnot efficiency. On the contrary, in the capacitive-osmotic cell in Figure 3(a), the charging voltage is varied in an isothermal process, by adjusting the electrolyte concentration ($c$). The system is immersed in a thermal reservoir. When the piston compresses the liquid phase, $c$ is increased and $V$ is lowered; vice versa. The piston operation consumes work ($W_{\mathrm{os}}$). In equilibrium, the second law (Equation 4) ensures that $W_{\mathrm{e}}=W_{\mathrm{os}}$ (Equation 6) and the net work produced is zero. With the intrinsic-nonequilibrium ion distribution, Equation (5) cannot be satisfied and, consequently, $W_{\mathrm{e}}>W_{\mathrm{os}}$.

Besides the osmotic pressure, there may be other concentration-dependent phenomena that can be utilized to demonstrate the SND effect. Section A4 of the Appendix shows one example, where a reference equilibrium cell is employed to reduce the charging voltage of the intrinsic-nonequilibrium cell. The setup does not contain any moving parts (e.g., the osmotic piston).

It is essential to distinguish the intrinsic-nonequilibrium steady state from a conventional nonequilibrium process. The latter is sustained by external fluxes of energy or mass; once the extrinsic driving forces are removed, the system would relax to equilibrium. In contrast, the intrinsic-nonequilibrium state shown in Figure 5 and [10-16] exhibits the following three characteristics simultaneously: First, the ion distribution is far from equilibrium (Figure 5c-e). Second, the out-of-equilibrium state is not maintained by any energy or mass flux with the environment. The system is isothermal and subject to no external driving force after the electrodes are charged to a given $Q$. Third, the nonequilibrium state is steady rather than transient or fluctuating (see Section 6.1). Even if initially in equilibrium, the effective surface ion density would spontaneously depart from equilibrium and evolve toward a different steady state. Section A1 of the Appendix provides a more general discussion.

## 6.4 Compatibility with the principle of maximum entropy

To understand intrinsic nonequilibrium, it is critical to explain how, without any external thermodynamic driving force, entropy does not increase to its maximum possible value [2-4]. The principle of maximum entropy reflects the basic logic that the macrostate with the highest probability (measured by entropy) is the most probable, which is "unconditionally" true.

In Figure 2(a), consider the dissolved ions as an isothermal system immersed in a thermal reservoir at a constant temperature $T$. The thermal reservoir includes the nanopore walls, the solvent (water), and the environment. For simplicity, assume that the microstates are discrete in the phase space. Entropy is defined as

$$S = -k_{\mathrm{B}} \sum_i f_i \ln f_i, \tag{13}$$

where $f_i$ is the probability of the $i$-th possible microstate of ion distribution. If the system is fully chaotic (e.g., if the pore size is much larger than the ion size), no detailed information is known about $f_i$, except for the following two constraints:

$$\sum_i f_i = 1, \tag{14}$$

$$\sum_i f_i \epsilon_i = U_{\mathrm{I}}, \tag{15}$$

where $\epsilon_i$ is the energy level of the $i$-th microstate, and $U_{\mathrm{I}}$ is the effective internal energy of the ions. Maximization of entropy requires that

$$\frac{\partial \mathcal{L}}{\partial f_i} = 0, \tag{16}$$

where $\mathcal{L} = -k_B f_i \ln f_i + \alpha_m(1 - \sum_i f_i) + \beta_m(U_I - \sum_i f_i \epsilon_i)$ is the Lagrangian, and $\alpha_m$ and $\beta_m$ are the Lagrange multipliers. The solution of Equation (16) is $f_i = e^{-(k_B + \alpha_m + \beta_m \epsilon_i)/k_B}$. Substituting it into Equation (14) leads to

$$f_i = \frac{1}{Z_A} e^{-\beta \epsilon_i}, \tag{17}$$

where $Z_A \triangleq e^{1+\alpha_m/k_B} = \sum_i e^{-\beta \epsilon_i}$ is the partition function, and $\beta = \beta_m/k_B = 1/(k_B T)$. Combination of Equations (17) and (13) gives the maximum possible entropy that the system can reach, i.e., the global maximum in the phase space in equilibrium:

$$S_{eq} = k_B \ln Z_A + k_B \beta U_I. \tag{18}$$

Any other $f_i$ (e.g., Equation 21 below) would result in a smaller entropy than $S_{eq}$.

When the effective nanopore size is slightly larger than the effective ion size ($d_i$) but less than $2d_i$, the ion movement in the nanoporous electrodes is confined in quasi-1D lineups and is less random than the chaotic case. More knowledge about the steady-state ion distribution ($\sigma^+$ and $\sigma^-$) is available, as discussed in Section 6.2. Specifically,

$$\sum_i f_i \sigma_i^+ = \sigma^+ \text{ and } \sum_i f_i \sigma_i^- = \sigma^-, \tag{19}$$

where $\sigma_i^+$ and $\sigma_i^-$ denote the effective surface densities of the confined counter-ions and co-ions of the $i$-th microstate, respectively. In Equation (19), the expectation values of surface ion density ($\sigma^+$ and $\sigma^-$) are treated as known parameters; for example, they may be calculated through MD simulations [10-16] or measured experimentally (Section 4). In an equilibrium system, $\sigma^+$ and $\sigma^-$ are proportional to the Boltzmann factor and Equation (19) is trivial, since it can be derived from Equation (17). For the small nanopores, as $\sigma^+$ and $\sigma^-$ are non-Boltzmannian, Equation (19) is nontrivial (i.e., it offers useful information) and should be taken into account in the Lagrangian:

$$\mathcal{L} = -k_B f_i \ln f_i + \alpha_m(1 - \sum_i f_i) + \beta_m(U_I - \sum_i f_i \epsilon_i) + \\ \theta_d(\sum_i f_i \sigma_i^+ - \sigma^+) + \theta_z(\sum_i f_i \sigma_i^- - \sigma^-), \tag{20}$$

where $\theta_d$ and $\theta_z$ are the additional Lagrange multipliers. The solution of Equation (16) becomes

$$f_i = \frac{1}{Z_B} e^{\phi_i} e^{-\beta \epsilon_i}, \tag{21}$$

where $\phi_i = (\theta_d \sigma_i^+ + \theta_z \sigma_i^-)/k_B$, and $Z_B = \sum_i e^{\phi_i} e^{-\beta \epsilon_i}$ is the generalized partition function.

To further examine $\phi_i$, based on Equation (8), define $V_\alpha = \lambda_V \ln(\sigma^+/\sigma_{ct})$ and $V_\beta = \lambda_V \ln(\sigma^-/\sigma_{co})$, where $\lambda_V = 2k_B T/e_z$ and $e_z = ze_0$. In terms of ion distribution, the confinement effect of small nanopores at $V$ is equivalent to the equilibrium state in large pores at $V_\alpha$ (for counter-ions) or $V_\beta$ (for co-ions). For such a reference large-pored equilibrium system,

$$f_i = f_\mathrm{r} e^{-\beta \bar{E}_i}, \tag{22}$$

where $f_\mathrm{r}$ is the probability of the baseline microstate with no confined ions, and $\bar{E}_i = \sigma_i^- V_\beta - \sigma_i^+ V_\alpha$. Equation (22) can be rewritten as $f_i \propto e^{\beta\theta_i} e^{-\beta\epsilon_i}$, where $\theta_i = \epsilon_i - \bar{E}_i = \sigma_i^+ \Delta V_\alpha - \sigma_i^- \Delta V_\beta$, $\Delta V_\alpha = V_\alpha - V$, and $\Delta V_\beta = V_\beta - V$. A comparison between $f_i \propto e^{\beta\theta_i} e^{-\beta\epsilon_i}$ and Equation (21) suggests that $Z_\mathrm{B} = 1/f_\mathrm{r}$ and $\phi_i = -\beta\theta_i$. Consequently, Equation (21) becomes

$$f_i = \frac{1}{Z_\mathrm{B}} e^{2\omega_i/e_\mathrm{z}} e^{-\beta\epsilon_i}, \tag{23}$$

where $\omega_i = \sigma_i^+ \ln(\sigma^+/\sigma_0^+) - \sigma_i^- \ln(\sigma^-/\sigma_0^-)$, and $\sigma_0^\mp = \sigma_\mathrm{n} e^{\mp\beta e_\mathrm{z} V/2}$. Based on the expressions of $\phi_i$ and $\omega_i$, $\theta_\mathrm{d} = \theta_0 \ln(\sigma^+/\sigma_0^+)$ and $\theta_\mathrm{z} = -\theta_0 \ln(\sigma^-/\sigma_0^-)$, where $\theta_0 = 2k_\mathrm{B}/e_\mathrm{z}$. If the ion distribution were in equilibrium, since $\omega_i = 0$, Equation (23) would be reduced to the Boltzmann factor. When $\sigma^+$ and $\sigma^-$ are intrinsically out of equilibrium, at a given energy level $\epsilon_i$, because of the nonchaoticity factor $e^{2\omega_i/e_\mathrm{z}}$, $f_i$ varies with $\sigma_i^+$ and $\sigma_i^-$, i.e., Equation (23) does not satisfy Boltzmann's assumption of equal *a priori* equilibrium probabilities.

Combination of Equations (23) and (11) gives the intrinsic-nonequilibrium entropy

$$S_\mathrm{ne} = k_\mathrm{B} \ln Z_\mathrm{B} + k_\mathrm{B}\beta U_\mathrm{I} - \theta_0 \Phi_\mathrm{I}, \tag{24}$$

where $\Phi_\mathrm{I} = \sum_i f_i \omega_i$. As the intrinsic-nonequilibrium $f_i$ in Equation (23) differs from the equilibrium $f_i$ in Equation (17), $S_\mathrm{ne} < S_\mathrm{eq}$.

On the one hand, $S_\mathrm{ne}$ is maximized, because it is obtained from Equation (16). On the other hand, with the additional restrictions on $f_i$ (Equation 19), Equation (23) reflects the more constrained maximization of $S$, such that $S_\mathrm{ne}$ is a local maximum in the phase space. Notice that $S_\mathrm{ne}$ is not associated with any metastable state. Without violating the microscopic governing equations, no accessible state can have a higher entropy than $S_\mathrm{ne}$, i.e., $S_\mathrm{ne}$ is the maximum possible entropy at the steady state ($S_\mathrm{Q}$). Since $S_\mathrm{ne}$ is a function of $V$, when $V$ varies, entropy can spontaneously decrease from a higher $S_\mathrm{ne}$ to a lower $S_\mathrm{ne}$.

The second law of thermodynamics may be generalized as follows [5]: in an isolated system, entropy cannot evolve away from $S_\mathrm{Q}$; that is,

$$S \to S_\mathrm{Q}. \tag{25}$$

If the system can relax to thermodynamic equilibrium, $S_{\mathrm{Q}} = S_{\mathrm{eq}}$ and Equation (25) is the same as the conventional statement of the second law. If the system reaches the intrinsic-nonequilibrium steady state, $S_{\mathrm{Q}} = S_{\mathrm{ne}} < S_{\mathrm{eq}}$. It is the root cause of the work production in Figure 3.

In the above analysis, the temperature field is assumed to be uniform. If we take into consideration the possible heterogeneous distribution of $\bar{K}$ [2-4], $S_{\mathrm{ne}}$ would be even smaller than Equation (24), as the degree of nonuniformity is higher.

### 6.5 Open questions and future research

The nanoporous electrode in Figure 2(a) has two sides. On the right-hand (outer) side of the surface zone, the ion behavior is unconstrained, following the Maxwell-Boltzmann distribution. On the left-hand (inner) side of the surface zone, the adsorbed ions are confined. The ion behavior at the inner side raises a number of open questions. For example, ion correlations [33], confinement-induced ion reorientation, and nonuniform surface-charge distributions may influence the ion adsorption capacity. In the small nanopores, the quasi-1D ion motion does not involve ordinary scattering effects. Compared to the heavy ions, the momenta of water molecules tend to be much smaller. These details may be analyzed through MD simulations or using modified Poisson-Boltzmann equations, to account for steric exclusion, ion polarizability, finite-size packing, etc. The assumption of local equilibrium needs to be carefully examined [2], especially on the small yet critical length scale of SND. Regardless of the specific mechanisms, the theoretical framework must explain the intrinsic-nonequilibrium nature of Figure 5.

There are limitations in our experiment and its data analysis. This study focuses on the steady-state ion distribution. The net heat exchange with the environment and the work extracted in an isothermal cycle are not directly measured. Consequently, Equation (6) is investigated theoretically. Equations (8) and (9) assume symmetry between co-ion and counter-ion. The evaluation of Figure 5(d) relies on macroscopic averages. The BET surface area is used as the upper bound; the micropore volume is approximate; the actual pore-size distribution is represented by its mean value. The resulting nonequilibrium term in Figure 5(e) therefore carries statistical uncertainty, although the order-of-magnitude discrepancy remains robust against reasonable variations in these quantities. Moreover, the microscopic picture of quasi-1D ion transmission is inferred from geometric constraints (Figure 2a) and published MD simulations [10-12]; no in situ measurements of axial profiles or transmission probabilities are carried out.

Future experiments may be conducted on wider ranges of sample size and energy levels. Direct characterization of ion occupancy and transit statistics within a single, well-characterized nanopore or nanotube would isolate the confinement contribution without the ensemble averaging. Extending the temperature window well beyond ambient conditions would enable quantification of the activation parameters governing axial ion transmission and help to distinguish its thermodynamic origin from the temperature-independent geometric factors. Parallel studies using ionic liquids with various ion sizes or micro-electrodes with functionalized surface chemistry can further map the boundary between equilibrium and intrinsic-nonequilibrium regimes.

Because of the large ion size and the low ion concentration, compared to regular double-layer supercapacitors, the CsPiv cells have a relatively low energy density. It is important to explore whether the concept of energy-barrier SND can be applied to other mesoscopic physical systems. For instance, the upper limit of the power density of a free-electron Fermi gas (e.g., the conduction electrons in a metal or graphene) may be more than 10 $kW/cm^3$ [2]. In Figure 3(a), the bulk liquid phase represents a low-energy state of the charge carriers, and the charged microporous electrode represents a high-energy state.

Besides the Coulomb force, other relevant thermodynamic forces include degeneracy pressure, chemical potential, magnetic forces, inertial and gravitational forces, and gas or plasma pressure. The unique thermal state associated with SND merits further study [4]. The implications of intrinsic nonequilibrium for cosmic evolution is an interesting direction for future research.

## 7. Concluding Remarks

It is well established that thermodynamics may not apply to certain small or single-particle systems. The present investigation experimentally demonstrates that a macroscopic many-particle system can also be non-thermodynamic and, more importantly, have nontrivial energetic consequences, e.g., the large $|\delta_V|$ measured in Figure 5(d). The steady-state distribution of large ions in charged small nanopores is *intrinsically* out of equilibrium, thereby breaking the boundaries of the second law of thermodynamics and opening the possibility of producing useful work in a cycle by absorbing heat from a single thermal reservoir with no other effect.

Such a counterintuitive phenomenon does not challenge the second law within its established domain of application (fully chaotic systems) but extends our understanding of thermodynamics beyond the traditional framework to locally nonchaotic systems. The experimental data are consistent with the published MD simulations [10-16]. The data analysis is compatible with the reference tests on small-ion cells [17].

The experiment is inspired by the recent theoretical study of the foundations of statistical mechanics [2], with the weak gravitational force in the "toy model" (Figure 1d) being replaced by a strong Coulomb force. The supercapacitive cells consist of nanoporous carbon electrodes immersed in dilute aqueous cesium pivalate (CsPiv) solutions. The key characteristic is that the effective nanopore size is slightly larger than the effective ion size ($d_{\mathrm{i}}$) but less than $2d_{\mathrm{i}}$, so that the motion of the confined ions is quasi-1D and less chaotic than that in large pores.

Sections 2-3 reaffirm that according to the second law, the steady-state ion distribution ($\sigma^+$ and $\sigma^-$) must be in equilibrium. An intrinsic-nonequilibrium $\sigma^+$ and $\sigma^-$ would lead to a mismatch between the chemical potential and the electric potential. Equation (5) represents the heat-engine statement of the second law; Equation (8) represents the entropy statement of the second law; Equation (11) states that Equations (5) and (8) should be equivalent. When the effective electrode surface charge density $\sigma$ is fixed (i.e., with a given state of charge), the cell potential ($V$) varies with the electrolyte concentration ($c$). The concentration sensitivity of electric potential ($\delta_V \triangleq \partial V/\partial c$) cannot be arbitrary: $\delta_V$ should obey not only the second law (Equation 5) but also charge balance (Equation 7). Only the equilibrium ion distribution (Equation 11) can meet both requirements.

Sections 4-5 show that, remarkably, the steady-state ion distribution in the CsPiv cell intrinsically differs from equilibrium, without an external driving force. The measured charge efficiency $\Lambda$ (Figure 5c) is nearly one order of magnitude smaller than the prediction of the second law (Equation 12). As a result, $|\delta_V|$ significantly exceeds the upper limit of conventional thermodynamic analysis (Figure 5d).

The confinement effect of the nanopore walls plays the role of a spontaneous-nonequilibrium domain (SND). A number of observations support that the measurement reflects the steady state (Section 6.1). In addition to accelerated diffusion and constrained relaxation of ion conformation, the mechanisms of intrinsic nonequilibrium may also involve surface heterogeneities and excess ion energy. Intrinsic nonequilibrium is not in conflict with the basic principle of maximum entropy.

**Acknowledgment:** Special thanks are due to Dr. Rui Kou, Dr. Daniel J. Noelle, Mr. Mingyuan Zhang, and Dr. Zhaoru Shang for their help with system design and data collection.

## Appendix

### A1. Intrinsic nonequilibrium

Intrinsic nonequilibrium is fundamentally distinct from conventional nonequilibrium phenomena. A regular nonequilibrium process is either sustained by energy or mass fluxes with the environment, or is transient or fluctuating [8]. If the external driving force is removed, the system eventually relaxes to equilibrium. In contrast, intrinsic nonequilibrium has the following three characteristics: First, the state is significantly different from equilibrium. Second, there is no external driving force to maintain the system away from equilibrium, such as a pressure or temperature gradient imposed by the environment. Third, the state is steady, neither transient nor fluctuating. Regardless of how long the system is allowed to evolve, equilibrium cannot be reached. The systems under investigation are closed and immersed in a thermal reservoir (or isolated). The mechanism underlying intrinsic nonequilibrium is rooted in inherent properties of the system, e.g., local nonchaoticity.

Whether intrinsic-nonequilibrium systems exist is a fundamental question in statistical mechanics. Both Maxwell and Boltzmann studied it [e.g., 44,45]. In fact, people have long been aware that there are small-sized intrinsic-nonequilibrium setups, such as certain nonchaotic or nonergodic particle movements. One simple example is a single elastic particle bouncing vertically in a gravitational field over a horizontal specular floor. The particle obeys Newton's second law, not the second law of thermodynamics. The probability ($\bar{p}_{\mathrm{z}}$) of finding the particle at height $z_{\mathrm{h}}$ is inversely proportional to the vertical component of particle velocity ($v_{\mathrm{z}}$), not proportional to the Boltzmann factor. Traditionally, these non-thermodynamic models are regarded as "unimportant", because they are microscopic in scale and their capacity for work production is trivial.

Based on Figure 1(a), the concept of Figure 1(b) is quite straightforward: We attach two large horizontal shelves at the top and bottom of the vertical plane, respectively. On the one hand, there are many particles on the large shelves, so that the system is macroscopic and its work-producing capacity may be nontrivial. On the other hand, the height difference ($z_{\mathrm{G}}$) between the two shelves is much smaller than the nominal particle mean free path ($\lambda_{\mathrm{F}}$), so that the local particle trajectories in the vertical plane are nonchaotic, and the intrinsic-nonequilibrium characteristics may be preserved.

In Figure 1(b), the low-height step serves as the spontaneous-nonequilibrium domain (SND). It is not a regular narrow band: inside the step, and only inside the step, does the external force field (gravity) take effect. The SND allows the intrinsic-nonequilibrium effects to "spread" from the small region (the low-height vertical step) to the surrounding large areas, leading to the significant overall energetic consequences.

It is worth noting that in the aforementioned single-particle model, $\bar{p}_{\mathrm{z}} \propto 1/v_{\mathrm{z}}$ does not satisfy Boltzmann's assumption of equal *a priori* equilibrium probabilities [1]. When the particle moves upward and its speed decreases, it spends a longer time at higher positions; vice versa. Consequently, although the system remains at the same energy level $E_{\mathrm{z}}$, for $z_{\mathrm{h}} < E_{\mathrm{z}}/(mg)$, the particle is more likely to be found at a higher position than at a lower position. The contradiction with Boltzmann's assumption does not imply that $\bar{p}_{\mathrm{z}} \propto 1/v_{\mathrm{z}}$ is incorrect, nor that thermodynamic analysis should never be applied to this type of problem. Instead, it suggests the need for a new theory.

It is also worth noting that the intrinsic-nonequilibrium systems studied here (the models in Figure 1 and the experiment in Section 4) are not perpetual motion machines of the second kind. A perpetual motion machine of the second kind violates the second law of thermodynamics. It is based on fully chaotic processes and employs equilibrium components, yet attempts to produce nonequilibrium effects without an energetic penalty. On the contrary, an intrinsic-nonequilibrium system contains locally nonchaotic SND, e.g., the low-height step in Figure 1(b) or the nanopores in Figure 2(a). The second law is not violated, in the sense that SND lies beyond the conventional framework of thermodynamics. The concept of intrinsic nonequilibrium broadens statistical mechanics by extending its boundaries to regimes where Boltzmann's H-theorem does not apply.

## A2. Cross-influence of thermodynamic forces

Consider a closed thermodynamic system immersed in a thermal reservoir. It has two thermally correlated thermodynamic forces, $F_1$ and $F_2$. Their conjugate variables are $x_1$ and $x_2$, respectively. Define the cross-influence of $F_1$ and $F_2$ as $\delta_{12} = \partial F_1/\partial x_2$ and $\delta_{21} = \partial F_2/\partial x_1$.

As shown in Figure 7(a), initially, at State I, $x_1$ is changed by an arbitrarily small amount $\mathrm{d}x_1$. Correspondingly, $F_2$ varies by $\delta_{21}\mathrm{d}x_1$. Then, $x_2$ is changed by an arbitrarily small amount $\mathrm{d}x_2$, and $F_1$ varies by $\delta_{12}\mathrm{d}x_2$. From State III to IV, $x_1$ changes back by $-\mathrm{d}x_1$. Finally, from State

IV to I, the system returns to the initial state. Without loss of generality, $\delta_{12}$ and $\delta_{21}$ are depicted as positive. In the isothermal cycle, $F_2$ produces work $W_{\mathrm{F2}} = \delta_{21}\mathrm{d}x_1\mathrm{d}x_2$, and $F_1$ consumes work $W_{\mathrm{F1}} = \delta_{12}\mathrm{d}x_2\mathrm{d}x_1$. The heat-engine statement of the second law demands that $W_{\mathrm{F2}} = W_{\mathrm{F1}}$. Thus, $\delta_{12} = \delta_{21}$, i.e., Equation (3). It may be viewed as a generalized Maxwell's relation.

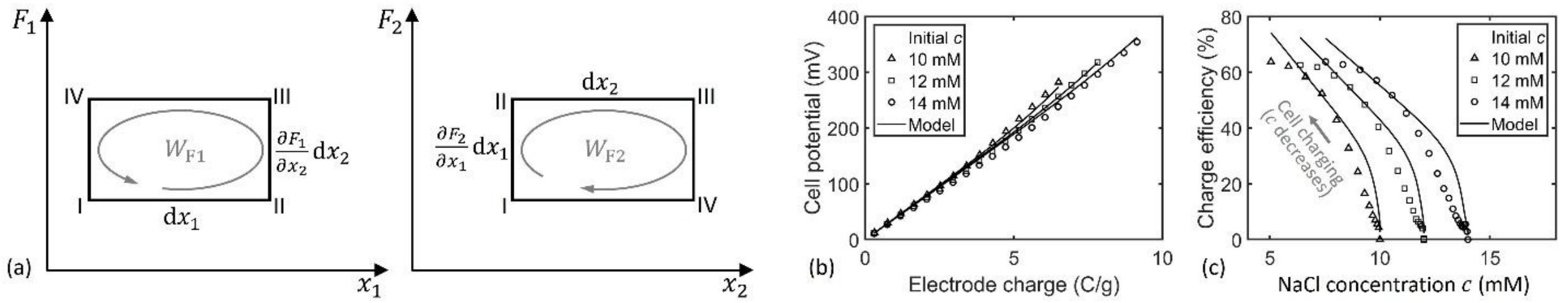

**Figure 7. (a)** A four-step isothermal cycle of a thermodynamic system. Indexes I-IV indicate the system states. The operation is reversible. The second law of thermodynamics demands that $W_{\mathrm{F1}} = W_{\mathrm{F2}}$. Experimental results of the reference NaCl cells [17]: **(b)** The charge curves. The cell potential ($V$) increases with the electrode charge ($Q$). The hollow symbols are experimental data; the solid curves are calculated from Equation (26). **(c)** The charge efficiency ($\Lambda$) as a function of the NaCl concentration ($c$). As the cell is charged, $c$ keeps decreasing and $\Lambda$ keeps increasing. The hollow symbols are experimental data; the solid curves are calculated from Equation (12). Panels (b) and (c) are adapted from [17].

A3. Reference test on NaCl cells

A similar experimental setup and procedure were used to test sodium chloride (NaCl) solutions with the same carbon electrodes; details have been published in [17]. The only major difference from the experiment in Section 4 is that the NaCl ions are much smaller than the CsPiv ions, such that $d_{\mathrm{e}} > 2d_{\mathrm{i}}$ for NaCl. The testing result is summarized in Figure 7(b,c), which fits well with the second law of thermodynamics (Equations 12 and 26).

Figure 7(b) shows typical charge curves with various initial NaCl concentrations ($c$). The initial $c$ is controlled by liquid replacement. As $c$ increases, with the same electrode charge ($Q$), the cell potential ($V$) is lower. In an equilibrium system, $\delta_\sigma = 0$. For symmetric counter-ion and co-ion, denote the equilibrium $\sigma_{\mathrm{ct}}$ and $\sigma_{\mathrm{co}}$ by $\bar{\sigma}_{\mathrm{n}}$, which is independent of $V$ ($b_{\mathrm{n}} = 1$). Equation (10) is reduced to $\delta_V = -\kappa_{\mathrm{p}}\sigma/\sqrt{\sigma^2 + 4\bar{\sigma}_{\mathrm{n}}^2}$ (for $z = 1$), where $\kappa_{\mathrm{p}} = 2k_{\mathrm{B}}T/(e_0 c)$. Integration of both sides yields

$$V = \frac{Q}{C_{\mathrm{s}}} + \frac{2k_{\mathrm{B}}T}{e_0}\ln\frac{2\bar{\sigma}_{\mathrm{n}}}{\sqrt{\sigma^2+4\bar{\sigma}_{\mathrm{n}}^2}-\sigma}, \tag{26}$$

where $Q/C_\mathrm{s}$ is the Stern-layer-like term. The solid curves in Figure 7(b) are calculated from Equation (26), where $\chi_0\alpha_\mathrm{e} = 2.9$ and $C_\mathrm{s} = 29.0$ F/g. These values are compatible with the literature data of nanoporous carbon adsorbents [38].

Figure 7(c) compares the experimental data of charge efficiency (Λ) with Equation (12). Equation (12) uses the same settings as Equation (26), without any additional adjustable parameters: $\chi_0\alpha_\mathrm{e}$ is determined by Figure 7(b), and $C_\mathrm{s}$ does not influence Λ. The horizontal axis is the real-time $c$. As the cell is charged, $c$ keeps decreasing and Λ keeps increasing. In contrast to the order-of-magnitude discrepancy observed for the CsPiv cells in Figure 5(c), the measurements on the NaCl cell are in reasonable agreement with conventional thermodynamic analysis. Near the end of the charging process, the experimental data deviate from the solid curves, likely due to ion starvation.

## A4. A two-cell model system

The ideal-gas model in Figure 1(e) is two-ended and asymmetric. One end (the wide ramp) is in equilibrium, and the other end (the low-height step) is intrinsically out of equilibrium. Partly inspired by this configuration, Figure 8(a) depicts a two-cell system, consisting of an intrinsic-nonequilibrium capacitive cell ("n") and a reference equilibrium cell ("r"). Cell "n" uses two small-pored electrodes, wherein the steady-state ion distribution is non-Boltzmannian. Cell "r" uses two large-pored electrodes, wherein the ion distribution is proportional to the Boltzmann factor. The two cells share the same electrolyte solution, connected through a salt bridge. The system is closed and immersed in a thermal reservoir.

Figure 8(b) shows an isothermal operation cycle. In this section, subscript "n" indicates the intrinsic-nonequilibrium cell, and subscript "r" indicates the reference equilibrium cell. From State I to II (switch "N1" is on; all the other switches are off), the intrinsic-nonequilibrium cell is charged. For simplicity, assume that the change of electrode charge ($\mathrm{d}Q_\mathrm{n}$) is arbitrarily small. The voltage of the equilibrium cell ($V_\mathrm{r}$) increases, as the electrolyte concentration is reduced. From State II to III (switch "R1" is on; all the other switches are off), the reference cell is charged by an arbitrarily small amount $\mathrm{d}Q_\mathrm{r}$. The voltage of the intrinsic-nonequilibrium cell ($V_\mathrm{n}$) increases, as the electrolyte concentration is reduced. From State III to IV (switch "N2" is on; all the other switches are off), the intrinsic-nonequilibrium cell is discharged by $\mathrm{d}Q_\mathrm{n}$, which reduces the voltage of the reference cell. From State IV to I (switch "R2" is on; all the other switches are off), the reference cell is discharged by $\mathrm{d}Q_\mathrm{r}$, which reduces the voltage of the intrinsic-nonequilibrium cell.

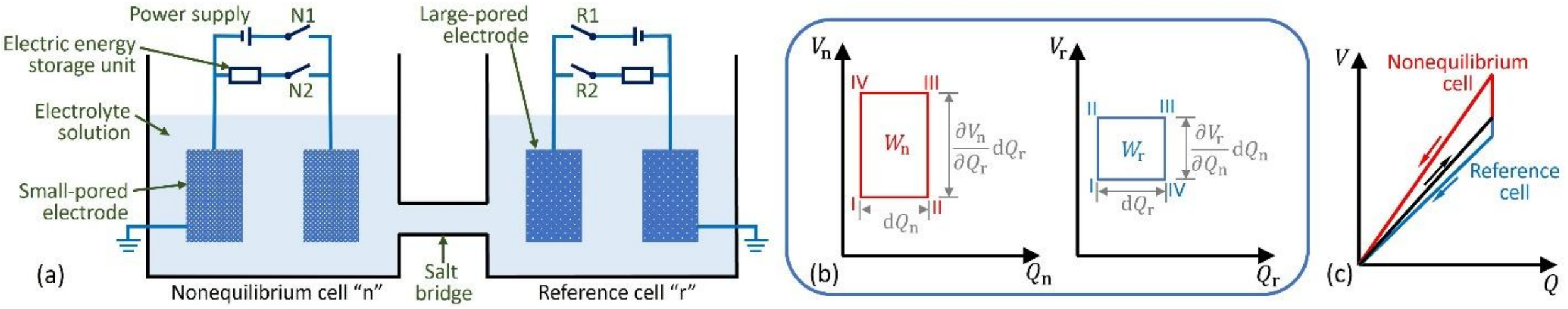

**Figure 8. (a)** Schematic of the two-cell system. It is closed and immersed in a thermal reservoir. **(b)** An isothermal cycle. Indexes I-IV indicate the system states. **(c)** Another isothermal cycle. The two cells have the same capacitance, such that their charge curves (the black line in the middle) are identical. The discharge curves (the red and blue lines) are different.

The intrinsic-nonequilibrium cell produces work $W_\mathrm{n}$, and the reference cell consumes work $W_\mathrm{r}$. The cross-influence of the two cells is achieved through ion diffusion across the salt bridge. As the reference cell is charged, its electrodes absorb ions. Hence, the electrolyte concentration in both cells ($c$) decreases, causing the electric potential of the intrinsic-nonequilibrium cell ($V_\mathrm{n}$) to increase by $(\partial V_\mathrm{n}/\partial Q_\mathrm{r})\mathrm{d}Q_\mathrm{r}$, and vice versa. From Figure 8(b), it can be seen that $W_\mathrm{n} = (\partial V_\mathrm{n}/\partial Q_\mathrm{r})\mathrm{d}Q_\mathrm{r}\mathrm{d}Q_\mathrm{n}$ and $W_\mathrm{r} = (\partial V_\mathrm{r}/\partial Q_\mathrm{n})\mathrm{d}Q_\mathrm{n}\mathrm{d}Q_\mathrm{r}$.

For the reference cell, $\partial V_\mathrm{r}/\partial Q_\mathrm{n} = (\partial V_\mathrm{r}/\partial c)(\partial c/\partial Q_\mathrm{n})$. Since the cell follows the second law of thermodynamics (Equation 5), $\partial V_\mathrm{r}/\partial c = -2k_\mathrm{B}T\Lambda_\mathrm{r}/e_0 c$, where $\Lambda_\mathrm{r} = \partial N_\mathrm{r}/\partial Q_\mathrm{r}$ and $N_\mathrm{r}$ are the charge efficiency and the adsorbed ions of the reference cell, respectively. Notice that $\partial c/\partial Q_\mathrm{n} = -V_\mathrm{p}^{-1}(\partial N_\mathrm{n}/\partial Q_\mathrm{n}) = -\Lambda_\mathrm{n}/V_\mathrm{p}$, where $V_\mathrm{p}$ is the volume of the liquid phase, and $\Lambda_\mathrm{n} = \partial N_\mathrm{n}/\partial Q_\mathrm{n}$ and $N_\mathrm{n}$ are the charge efficiency and the adsorbed ions of the intrinsic-nonequilibrium cell, respectively. Hence, $\partial V_\mathrm{r}/\partial Q_\mathrm{n} = 2k_\mathrm{B}T\Lambda_\mathrm{r}\Lambda_\mathrm{n}/(V_\mathrm{p}e_0 c)$.

For the intrinsic-nonequilibrium cell, $\partial V_\mathrm{n}/\partial Q_\mathrm{r} = (\partial V_\mathrm{n}/\partial c)(\partial c/\partial Q_\mathrm{r})$. Like the reference cell, $\partial c/\partial Q_\mathrm{r} = -\Lambda_\mathrm{r}/V_\mathrm{p}$. However, unlike the reference cell, because the adsorbed ion concentration is inherently out of equilibrium, Equation (5) cannot be satisfied, i.e., $\partial V_\mathrm{n}/\partial c \neq -2k_\mathrm{B}T\Lambda_\mathrm{n}/(e_0 c)$ (Figure 5d). Therefore, $\partial V_\mathrm{n}/\partial Q_\mathrm{r} \neq 2k_\mathrm{B}T\Lambda_\mathrm{n}\Lambda_\mathrm{r}/(V_\mathrm{p}e_0 c)$.

As $\partial V_\mathrm{n}/\partial Q_\mathrm{r} \neq \partial V_\mathrm{r}/\partial Q_\mathrm{n}$, $W_\mathrm{n}$ is different from $W_\mathrm{r}$. Specifically, the experimental data in Figure 5(d) suggest that $\partial V_\mathrm{n}/\partial Q_\mathrm{r}$ tends to be larger than its equilibrium counterpart $\partial V_\mathrm{r}/\partial Q_\mathrm{n}$. Therefore, $W_\mathrm{n} > W_\mathrm{r}$. In the four-step isothermal cycle in Figure 8(b), the system produces useful work $\Delta W = W_\mathrm{n} - W_\mathrm{r}$ by absorbing heat from the environment.

The two cells may be operated through other methods. One example is given in Figure 8(c). There is a valve in the salt bridge. Initially, the valve is open, and the electrolyte concentration ($c$) is equal on both sides. Then, the valve is closed, and switches N1 and R1 are turned on (switches N2 and R2 remain off). For simplicity, we assume that the charge curves of the two cells are identical, i.e., the two cells have the same capacitance. After they are charged, all the switches are turned off, and the salt-bridge valve is opened. During charging, because of the difference in charge efficiency, the small-pored electrodes absorb fewer ions and hence the intrinsic-nonequilibrium cell has a higher $c$. Across the salt bridge, ions diffuse to the reference cell. As $c$ decreases in the intrinsic-nonequilibrium cell, the cell potential increases. Correspondingly, in the reference cell, $c$ increases and the cell potential decreases. As the new steady state is reached, the valve is closed, and switches N2 and R2 are turned on to discharge the cells (switches N1 and R1 remain off). Finally, all the switches are turned off and the salt bridge is kept open, and the system returns to the initial state. Overall, in the isothermal cycle, the intrinsic-nonequilibrium cell generates electrical energy ($W_{\mathrm{n}}$), and the reference cell consumes electrical energy ($W_{\mathrm{r}}$). As shown in Figure 5(d), the intrinsic-nonequilibrium cell tends to have a higher $|\delta_V|$ than the reference cell. Therefore, $\Delta W = W_{\mathrm{n}} - W_{\mathrm{r}}$ is positive. The ion flow and $\Delta W$ represent the useful work produced through heat absorption from the environment.